\documentclass[journal]{IEEEtran}

\IEEEoverridecommandlockouts
\usepackage{bm}
\usepackage{cite}
\usepackage{amsmath,amssymb,amsfonts,mathrsfs}
\usepackage{algorithmic, booktabs}
\usepackage{graphicx}
\usepackage{graphics}
\usepackage{textcomp}
\usepackage[dvipsnames]{xcolor}
\usepackage{algorithm,algorithmic}[1]
\usepackage{setspace}
\usepackage{bbm}
\usepackage[bb=boondox]{mathalfa}

\usepackage{blindtext}
\usepackage{caption}
\usepackage{subcaption}
\usepackage{nomencl}
\usepackage{siunitx}
\usepackage[export]{adjustbox}
\def\BibTeX{{\rm B\kern-.05em{\sc i\kern-.025em b}\kern-.08em
T\kern-.1667em\lower.7ex\hbox{E}\kern-.125emX}}
\usepackage{etoolbox}
\usepackage{acronym}
\usepackage{multirow}
\usepackage{makecell}

\usepackage{nomencl}
\usepackage{multirow}
\usepackage{array}
\usepackage{pdflscape}
\usepackage{rotating}
\renewcommand\nomgroup[1]{%
  \item[\bfseries
  \ifstrequal{#1}{A}{Grid Connected Device Sets}{%
  \ifstrequal{#1}{B}{Grid Connected Device Parameters}{%
  \ifstrequal{#1}{C}{Grid Connected Device Variables}{%
  \ifstrequal{#1}{D}{Opt. Parameters, Variables \& Sets}{%
  \ifstrequal{#1}{E}{Mathematical Operations}{}}}}}%
]}

\allowdisplaybreaks

\begin{document}

\title{Optimal SVI-Weighted PSPS Decisions with Decision-Dependent Outage Uncertainty}


\author{Ryan Greenough,
Kohei Murakami, Jan Kleissl, and Adil Khurram 
	\thanks{Ryan Greenough, Kohei Murakami, and \hbox{Adil Khurram} are with the Department
of Mechanical and Aerospace Engineering, University of California at San
Diego, La Jolla, CA 92093 USA (e-mail: rgreenou@ucsd.edu; k1murakami@ucsd.edu; akhurram@ucsd.edu)} 
\thanks{
Jan Kleissl is with the Center for Energy Research, Department of
Mechanical and Aerospace Engineering, University of California at San
Diego, La Jolla, CA 92093 USA (e-mail: jkleissl@ucsd.edu).}}


\maketitle
\makenomenclature

\begin{abstract}

Public Safety Power Shutoffs (PSPS) are a pre-emptive strategy to mitigate the wildfires caused by power system malfunction. System operators implement PSPS to balance wildfire mitigation efforts through de-energization of transmission lines against the risk of widespread blackouts from load shedding. 

Existing approaches do not incorporate decision-dependent wildfire-driven failure probabilities, as modeling outage scenario probabilities requires incorporating high-order polynomial terms in the objective. This paper uses distribution shaping to develop an efficient MILP problem representation of the distributionally robust PSPS problem. The wildfire risk of operating a transmission line is a function of the probability of a wildfire-driven outage and the acres burned from that transmission line if it causes a wildfire. 

A day-ahead unit commitment and line de-energization PSPS framework is used to assess the trade-off between total cost and wildfire risk at different levels of distributional robustness, parameterized by a level of distributional dissimilarity $\kappa$. IEEE RTS 24-bus test system simulations show that allowing distributional dissimilarity can reduce monthly averaged reduced out-of-sample costs by 1\% when compared to the risk-neutral approach. 

\end{abstract}


\vspace{-2.5em}
\mbox{}
\nomenclature[]{\(\mathcal{B},\mathcal{N}_i\)}{Set of buses, Set of neighboring buses to bus $i$}
\nomenclature[]{\(\mathcal{L}\)}{Set of transmission lines (edges)}
\nomenclature[]{\(\mathcal{K}\)}{Set of damaged lines (edges)}
\nomenclature[]{\(\mathcal{U}\)}{Undirected graph model of power grid}
\nomenclature[]{\(H\)}{Number of samples in DA optimization horizon}
\nomenclature[]{\(B_{\ell}\)}{System susceptance  value of line $\ell$}
\nomenclature[]{\(\theta_{i,t,s}\)}{Phasor angle at bus $i \in \mathcal{B}$ at time $t$, and scenario $s$}
\nomenclature[]{\(\underline{p}_{g},\overline{p}_{g}\)}{Lower/upper generation limit for each generator $g \in \mathcal{G}$}
\nomenclature[]{\(\underline{U}_{g},\overline{U}_{g}\)}{Lower/upper generation ramp limit for each $g \in \mathcal{G}$}
\nomenclature[]{\(p_{g,t,s}\)}{Active generation provided by each generator $g \in \mathcal{G}$ at time $t$ and scenario $s$}
\nomenclature[]{\(p_{d,t,\omega}\)}{Demand $d \in \mathcal{D}$, at time $t$, for demand scenario $\omega$ }
\nomenclature[]{\(C^{\mathrm{VoLL}}_{d}\)}{Value of Lost Load for each demand node $d \in \mathcal{D}$}
\nomenclature[]{\(x_{d,t,s}\)}{Fraction of demand $d \in \mathcal{D}$, time $t$, and scenario $s$}
\nomenclature[]{\(\mathcal{H}\)}{Set of time-steps in the optimization horizon}
\nomenclature[]{\(\mathcal{G}\)}{Set of generators}
\nomenclature[]{\(\mathcal{D}\)}{Set of demands}
\nomenclature[]{\(R_{\ell,t}\)}{Wildfire risk of a transmission line, $\ell$}
\nomenclature[]{\(R_{\text{Tot}}\)}{Cumulative system transmission line wildfire risk}
\nomenclature[]{\(z_{g,t},z_{\ell}\)}{Active status for generators and transmission lines}
\nomenclature[]{\(\hat{z}_{\ell}\)}{Truncated version of the active status for transmission line $z_{\ell}$ for line $\ell$.}
\nomenclature[]{\(p^{\text{aux}}_{g,t,s}\)}{Auxiliary variable for a generator $g \in \mathcal{G}$, time $t$, and scenario $s$ to prevent violation ramping constraints when generator $g$ is turned off}
\nomenclature[]{\(z^{\text{up}}_{g,t},z^{\text{dn}}_{g,t}\)}{Startup and shutdown binary decisions}
\nomenclature[]{\(z^{\text{dn}}_{\ell}\)}{De-energization of line $\ell$}
\nomenclature[]{\(t^{\text{MinUP}}_{g},t^{\text{MinDown}}_{g}\)}{Minimum up and down times for each $g \in \mathcal{G}$}
\nomenclature[]{\(\underline{\theta},\overline{\theta}\)}{Lower/upper limit on the phase angle difference}
\nomenclature[]{\(p_{\ell,t,s}\)}{Active power flow of line $\ell$, time $t$, and scenario $s$}
\nomenclature[]{\(\underline{p}_{\ell},\overline{p}_{\ell}\)}{Lower/upper thermal limits of line flow along $\ell$}
\nomenclature[]{\(\omega\)}{A demand scenario in a given set of scenarios $\Omega$}
\nomenclature[]{\(\Omega,\Omega^{\text{RT}}\)}{Set of day-ahead and real-time demand, WIP scenarios}
\nomenclature[]{\(\pi_{\omega}\)}{Probability of each demand and wildfire scenario $\omega$}
\nomenclature[]{\(\pi_{\ell,s}\)}{Probability of fire ignition near a line $\ell$ in scenario $s$}
\nomenclature[]{\(\mathbb{P}(\bm{z^{-}_{ij}})\)}{Line survival scenario probability distribution dependent on shut-off decision vector $\bm{z^{-}_{ij}}$}
\nomenclature[]{\(\mathcal{P}\)}{Ambiguity set for distribuionally robust optimization}
\nomenclature[]{\(\mathbb{Q}\)}{Worst-case line survival probability distribution}
\nomenclature[]{\(C_g\)}{Variable cost of generation for each $g \in \mathcal{G}$}
\nomenclature[]{\(C_{g}^{\mathrm{dn}},C_{g}^{\mathrm{up}}\)}{Cost to shutdown/startup a generator $g \in \mathcal{G}$}
\nomenclature[]{\(\mathcal{A}^{c}\)}{Complement of set $\mathcal{A}$}
\nomenclature[]{\(\|\cdot\|\)}{Cardinality of Set}
\nomenclature[]{\(f^{\mathrm{uc}},f^{\mathrm{oc}},f^{\mathrm{VoLL}}\)}{Unit commit (uc), operating (oc), VoLL costs}
\nomenclature[]{\(R_{\mathrm{tol}}\)}{Operator risk tolerance level }
\nomenclature[]{\(I_{\ell}\)}{Impact level (in acres burned or SVI) of a new fire started by line $\ell$}
\nomenclature[]{\(\iota_{\ell}\)}{Social Vulnerability Index weight of census tracts intersecting transmission line $\ell$}
\nomenclature[]{\(\Pi_s, \Phi_{s}\)}{Total/operating costs for a given outage scenario $s$ }
\nomenclature[]{\(\xi_{\ell,s}\)}{Bernoulli random survival status of line $\ell$, scenario $s$}
\nomenclature[]{\(\kappa\)}{Radius of ambiguity set}
\nomenclature[]{\(\gamma_{ms}\)}{Joint distribution between scenario $\bm{\xi}_{\bm{\ell},m}$ from $\mathbb{Q}$ and $\bm{\hat{\xi}}_{\bm{\ell},s}$ from $\mathbb{P}$}
\nomenclature[]{\(\delta_{ms}\)}{Measure of dissimilarity between scenario $\bm{\xi}_{\bm{\ell},m}$ from $\mathbb{Q}$ and $\bm{\hat{\xi}}_{\bm{\ell},s}$ from $\mathbb{P}$ }
\nomenclature[]{\(\tau\)}{Dual variable on the total variation constraint}
\nomenclature[]{\(v_s\)}{Maximum between the operational cost of scenario $s$ ($\Phi_s$) and the  operational cost Value at Risk ($\text{VaR}_{\kappa}(\bm{\Phi})$)}
\nomenclature[]{\(w_{s,j}\)}{Auxiliary variable for the product of $\pi_{L,s}$ and $\beta_j$ }
\nomenclature[]{\(\beta_{j}\)}{Binary variable equally one in scenario $j$ if the operational cost scenario $\Phi_{j}$ is the Value at Risk of the uncertain operational cost vector $\bm{\Phi}$}
\nomenclature[]{\(\hat{\pi}_{L,s}\)}{An approximation of decision dependent scenario probability $\pi_s(\bm{z^{-}_{\ell}})$ via distribution shaping}
\nomenclature[]{\(q_m\)}{Scenario probability of scenario $m$ from the candidate distribution $\mathbb{Q}$}
\nomenclature[]{\(\mathcal{S}\)}{Set containing all line outage scenarios}
\nomenclature[]{\(\mathcal{F}\)}{Power set of all line outage scenarios}

\printnomenclature

\section{Introduction}

While power system equipment failures account for less than 10\% of reported wildfire ignitions in California \cite{CPUC}, they are responsible for nine of the state's twenty most destructive wildfires \cite{Isaacs-Thomas, White}. To mitigate wildfire risk while maintaining grid reliability, electric utilities implement Public Safety Power Shutoffs (PSPS) \cite{Arab}. By temporarily de-energizing power lines in fire-prone areas, PSPS reduces the likelihood of utility-caused ignitions while ensuring safe and stable grid operation during the wildfire season \cite{CPUC, Arab}.

A fundamental challenge in PSPS planning is balancing wildfire risk mitigation through de-energization with minimizing load shedding. Traditional $N-1$ security-constrained approaches, which consider single-component failures, are often inadequate for PSPS since multiple power lines may intersect an active wildfire region. $N-k$ contingency-based approaches are more suitable but introduce significant computational challenges, as the feasibility set grows combinatorially with $k$, the number of de-energized components. Researchers have proposed less conservative PSPS planning strategies that weight $N-k$ planning strategies with the risk of transmission line-ignited wildfires using the U.S. Geological Survey’s (USGS) Wildfire Potential Index (WFPI) forecasts \cite{Rhodes, Kody, Rhodes2, Kody2, Taylor, Greenough, Piansky2}. These approaches incorporate wildfire-related grid impacts as exogenous inputs, affecting optimization parameters such as wildfire damage costs, transmission line outage probabilities, and thermal limits on power flow.

While some studies have formulated PSPS as a two-stage stochastic programming problem \cite{Greenough, Piansky} to incorporate the pre-emptiveness of decision-making and operational uncertainties, these approaches primarily hedge against demand uncertainty rather than wildfire uncertainty. Many optimization models \cite{Trakas2, Bayani, Rhodes, Kody, Rhodes2, Kody2, Taylor, Greenough, Piansky, Piansky2} rely on deterministic wildfire risk forecasts. \cite{Trakas2, Bayani, Rhodes, Kody, Rhodes2, Kody2, Taylor, Greenough, Piansky, Piansky2} neglect the uncertainty in wildfire risk while optimizing PSPS decisions by assuming perfect foresight of wildfire impact and ignition. However, in reality, simultaneous ignitions from multiple lines are rare; historically, occurring only five times in the last 30 years within the same region \cite{Westerling2}. In practice, assuming that all nonzero-risk transmission lines ignite leads to more conservative de-energization decisions and higher optimal costs.

Other approaches attempt to incorporate wildfire ignition uncertainty through scenario-based approximations \cite{Umunnakwe, Bayani2, Yang, Yang2, Greenough2}. Despite constructing probability distributions for wildfire ignition scenarios, \cite{Umunnakwe, Greenough2, Yang, Yang2} assume that the probability distribution is perfectly known. However, wildfires are low probable and high-impact events and their likelihood is inherently difficult to estimate. Furthermore, the probability distribution of wildfire ignitions may even shift due to climate change, making estimation more challenging.

\cite{Bayani2} employs a two-stage optimization for capacity expansion. However, the first-stage expansion planning strategy involves a Wait-and-See approach (or perfect foresight into each demand, renewable availability, and wildfire risk scenario). The second-stage operational decisions are assumed to be under the worst-case uncertainty which is more conservative than risk-averse second-stage operational decisions. 

\cite{Yang,Yang2} use two-stage and multi-stage stochastic programs to simulate operational PSPS decisions--excluding unit commitment-- during a wildfire event. However, the two-stage stochastic mixed integer program (SMIP) and multi-stage mixed integer program (M-SMIP) assume wildfire outage scenario probabilities are known, constant throughout the simulation, and independent of de-energization or other operational power flow decisions. The SMIP and M-SMIP assign the flammability of vegetation, measured by the WFPI, to each transmission line. Then \cite{Yang,Yang2} use the WFPI to adjust the relative wildfire ignition probabilities (WIP), which is shown in \cite{Greenough2} to lead to more costly under-commitment of generating resources due to under-estimating the severity of wildfire ignition events. \cite{Greenough2} creates a robust PSPS optimization that defines wildfire risk in terms of joint outage probabilities rather than in terms of impact based on WFPI. Furthermore, \cite{Greenough2} shows how the large fire probability (WLFP) maps better to WIP than WFPI leading to improved cost savings in a post-optimization real-time test. However, \cite{Greenough2} also assumes the probability of wildfire-driven outages from transmission lines is known and independent of operational decisions, such as transmission line power flows and day-ahead transmission de-energization decisions. \cite{Yang,Yang2,Greenough2} do not give operators the ability to tune optimization models for distributional uncertainty in wildfire ignition probability (WIP) nor introduce a feedback loop where system operators' de-energization decisions influence risk levels.

\cite{Moreira, Pianco} model how the outage probability dramatically increases when the line operates at its maximum flow rate compared to flowing power at less than the maximum flow rate. They explore how the maximum flow rate could alter outage likelihoods in a distributionally robust (DR) optimization for PSPS. More specifically, the ambiguity set accounting for the probability of each transmission line outage depends on the operational power flow decisions. However, this first-order moment-based uncertainty set constructs individual DR chance constraints for each line. Individual DR chance constraints are more conservative than considering a joint DR chance constraint capturing the joint distribution of different permutations of line outages because the joint chance constraints ensure all constraints are met simultaneously. The authors' previous work \cite{Greenough2} considers a joint chance constraint on the allowable probability of outages; however, the distribution is trimmed to include only the most probable damage outcome among a set of k-damaged lines. The optimal cost outcome is more risk-seeking because the less likely damage outcomes could be more damaging scenarios than the more probable scenarios. 

\cite{Pollack, Taylor} emphasize the socioeconomic impact of each line energization decision by scaling the wildfire risk of each energized transmission line by the average Social Vulnerability Index (SVI) of all census tracts that intersect the line. However, \cite{Pollack, Taylor} use a deterministic Optimal Power Shut-off (OPS) model and assume perfect foresight of the transmission line wildfire risks in units of WFPI rather than subsequent size or number of damaged structures. Not only would a two-stage stochastic model lead to less conservative operation than the deterministic counterpart (as shown in \cite{Greenough}) but quantifying the wildfire's impact in subsequent size or damaged structures may lead to better mapping of a wildfire's impact to the power delivery to specific communities than a generic wildfire metric.

In this work, we use WIP forecasts to estimate the nominal probability of wildfires produced from a transmission line outage and then optimize expected costs with respect to the worst-case candidate distribution that has some level of dissimilarity from the nominal distribution. We assume operators make decisions based on economic costs and use a tolerance level, $R_{\textrm{tol}}$, for an allowable cumulative impact of wildfires from energized lines. Then, we transform the Stochastic PSPS (SPSPS) proposed in \cite{Greenough} into a distributionally robust optimization with decision-dependent uncertainty DR-PSPS. In contrast to the authors' earlier works, \cite{Greenough} and \cite{Greenough2}, we model the wildfire ignitions as Bernoulli random events and a wildfire's subsequent impact in terms of acres burned. Then we compare costs between the SPSPS and DR-PSPS. And similar to \cite{Trakas2, Kody, Rhodes2, Kody2, Greenough, Umunnakwe, Yang, Yang2, Bayani, Bayani2, Piansky, Pollack, Piansky2, Taylor}, we only account for the wildfire risk of operating transmission lines.   

The optimization is solved in two stages for the IEEE RTS 24-bus transmission grid. In the day-ahead, unit commitment decisions are optimized based on the inputs of expected total demand at each bus and the WIP forecast for each line with respect to the probability of representative day-long scenarios. 
Then day-ahead unit commitment decisions from the SPSPS and DR-PSPS are tested on multiple line damage scenarios via a Monte Carlo simulation to represent real-time operation. Within the Monte Carlo simulation, we use real-time demand and a collection of transmission line outage scenarios based on the real-time WIP forecast near transmission lines. We analyze cost trends as the distributional robustness parameter $\kappa$ increases, highlighting the trade-off between conservatism and economic efficiency. 

The main contributions of this paper are the following: 
\begin{enumerate}

\item To the best of our knowledge, we propose the first two-stage decision-dependent DR unit commitment for PSPS (DR-PSPS) that explicitly models de-energization decision-dependent uncertainties in joint wildfire ignition probabilities and their subsequent impacts.

\item We extend the authors' day-ahead unit commitment model proposed in \cite{Greenough} and \cite{Greenough2} by modeling wildfire risk in both ignition probability and subsequent spread as well as incorporating uncertainty in wildfire ignition probabilities. 

\item We explore wildfire impacts in terms of cumulative Social Vulnerable Indices (SVIs) and acres burned. This characterization of a wildfire impact is more practical for operators than defining risk in unitless measures such as cumulative WFPI from \cite{Rhodes, Kody, Rhodes2, Kody2, Greenough, Bayani, Bayani2, Taylor, Piansky,Piansky2,Pollack} or normalized wildfire risk scores \cite{Bayani, Bayani2}.  

\end{enumerate}

The paper is organized as follows. Section~\ref{section:Modeling of Public Safety Power Shut-offs for Wildfire Risk Mitigation} introduces all elements of the decision-dependent stochastic PSPS and DR PSPS strategies including the day-ahead formulation and real-time optimization test. The stochastic optimization from the authors' prior work, \cite{Greenough}, is modified to incorporate endogenous uncertainty in the outage scenarios in Section~\ref{subsection:Constraint for the allowable wildfire risk considering both the uncertain wildfire instance, xi, and impact, I} and  
robustness to uncertainty in the probabilities of the outage scenarios in Section~\ref{subsubsection:Distributionally Robust Day-ahead PSPS formulation}. Section~\ref{section:Results} explains the results of the unit commitment on the IEEE RTS 24-bus system. Section~\ref{section:Conclusion} concludes the paper.

\section{Modeling PSPS for Wildfire Risk Mitigation}
\label{section:Modeling of Public Safety Power Shut-offs for Wildfire Risk Mitigation}


\subsection{Preliminaries}
\label{subsection:Preliminaries}
Let $\mathcal{U}=(\mathcal{B}, \mathcal{L})$ be the graph describing the power grid where $\mathcal{B}=\{1, \dots, B\}$ is the set of $B$ buses in the network, and $\mathcal{L}$ is the set of edges such that two buses $i,j \in \mathcal{B}$ are connected by a transmission line if $(i,j)\in \mathcal{L}$. The set of buses with generators and loads are collected in $\mathcal{G}$ and $\mathcal{D}$ respectively, and $\mathcal{H}=\{1, \dots, H\}$ is the set of time indices over the horizon of length $H$ of the optimization problem. A direct current approximation of the non-convex alternating current optimal power flow, called the DC-OPF is used to approximate the line power flows and bus power injections; all references to power are to its active power. Since the proposed DR-PSPS problem is solved using a scenario based approach, therefore, at timestep $t \in \mathcal{H}$, the power injected by the generator $g \in \mathcal{G}_i$ at bus $i \in \mathcal{B}$ in scenario $s \in \mathcal{S}$ is denoted by $p_{g,t,s}$. Similarly, $p_{d,t,s}$ is the load $d \in \mathcal{D}_i$ at bus $i \in \mathcal{B}$. 
Power flowing through the line $\ell \in \mathcal{L}$ is $p_{\ell,t,s}$ and the phase angle at bus $i\in \mathcal{B}$ is denoted by $\theta_{i,t,s}$. Finally, the binary variables are denoted by $z \in \{0,1\}$ with an appropriate subscript to capture component shut-off decisions. For the distribution shaping formulation in Section~\ref{subsubsection:Distribution Shaping for Second Stage Objective}, the edges $(i,j) \in \mathcal{L}$ are ordered and assigned a unique identifier $\ell_{ij} \in \overline{\mathcal{L}}$ where $\overline{\mathcal{L}} = \{ 1  \ldots L| \ell_{ij}=(i,j)\}$ so that there is a one to one mapping between $\mathcal{L}$ and $\overline{\mathcal{L}}$. We drop the $ij$ subscript from $\ell_{ij}$ to simplify the notation. However, the vector of all de-energization or line probabilities is $\bm{z^{-}_{ij}}$ and $\bm{\pi_{ij}}$ to avoid indexing confusion when creating truncated versions of $\bm{z^{-}_{ij}}$ called $\bm{\check{z}_{ij,\ell}}$. There are ordered sets of bus neighbors for each bus: $\mathcal{N}_i$.

Due to the scenario uncertainty, the objective function is stochastic, therefore, the SPSPS from \cite{Greenough} aims to minimize the expected economic costs, $\Pi_{s}$, across all scenarios $s \in \mathcal{S}$. In this paper, we expand on the model from \cite{Greenough} by letting the shut-off decisions influence the scenario probabilities $\pi_s$ and optimizing over a family of scenario probability distributions.

\subsection{Constraint for the allowable wildfire risk considering both the uncertain wildfire instance, $\bm{\xi_{ij}}$, and impact $\bm{I_{ij}}$}
\label{subsection:Constraint for the allowable wildfire risk considering both the uncertain wildfire instance, xi, and impact, I}

The wildfire risk of each transmission line ($R_{\ell}$) is modeled as a product of two components: WIP ($\pi_{\ell}$) and its subsequent impact ($I_{\ell}$). In the authors' prior works \cite{Greenough} and \cite{Greenough2}, \cite{Greenough} considered wildfire impact with a unitless measure of vegetation flammability called WFPI, and~\cite{Greenough2} only considers WIP. The new model developed in this work requires both a measure of WIP and wildfire impact.

Defining $\mathcal{S} = \{0,1\}^L$ as the finite sample space, the line survival scenarios can be denoted as $\bm{\xi_{ij,S}} = \biggl\{\bm{\xi}_{\bm{ij},1}, \bm{\xi}_{\bm{ij},2}, \ldots, \bm{\xi}_{\bm{ij},S}\biggr\}=\mathcal{S}$. The set of scenarios, $\mathcal{S}$, corresponds to the $2^L$ permutations of an $L$ dimensional vector representing the wildfire damage to each line $(i,j) \in \mathcal{L}$ where $L$ is the total number of lines. Let $\mathcal{F}$ be the power set of $\mathcal{S}$ and $\mathbb{P}$ be the probability measure, and $(\mathcal{S}, \mathcal{F}, \mathbb{P})$ the probability space. 

With the sample space, the random status that a wildfire ignition is caused by the transmission line $\ell$ in scenario $s$ is represented by $\xi_{\ell,s}$, and the WIP is represented by $\pi_{\ell}$. Without the influence of $\bm{z^{-}_{\ell}}$, $\mathbb{P}(\xi_{\ell,s}=0) = \pi_{\ell}$ and $\mathbb{P}(\xi_{\ell,s}=1) = 1-\pi_{\ell}$. The wildfire risk $R_{\ell}$ of an energized transmission line is defined as a product of the operator's decision to keep the line active, $z_{\ell}$, a Bernoulli random transmission line survival status, $\xi_{\ell}$, and the subsequent socioeconomic weighted impact of the resulting fire, $\tilde{I}_{\ell}=\iota_{\ell}I_{\ell}$. Socioeconomic weight $\iota_{\ell}$ is chosen as the SVI~\cite{SVI}. The fire's subsequent impact, $I_{\ell}$ is measured in burned acres or damaged structures. Therefore, the total wildfire risk on the network is the following,
\begin{align}
    R_{\text{tot}}=\sum_{\ell \in \mathcal{L}}z_{\ell}R_{\ell}=\sum_{\ell \in \mathcal{L}}z_{\ell}(1-\xi_{\ell})\tilde{I}_{\ell}.
    \label{eq:R_tot}
\end{align}
We make a further simplification and consider the worst-case risk $R_{\ell}=\pi_{\ell}\mathbb{E}[\tilde{I_{\ell}}]$.

Constraint~\eqref{eq:WFPI_Nmk} from the authors' earlier work (\cite{Greenough}) represents a line shut-off strategy that constrains the total number of active lines so that the sum of each line risk, $R_{\ell,t}$, does not exceed $R_{\text{tol}}$, which is an operator-selected parameter. $R_{\text{tol}}$ guarantees a certain level of security for the system while also penalizing the operation of transmission lines within regions of higher wildfire risk. 
\begin{align} \sum_{\ell \in \mathcal{L}} z_{\ell,t}R_{\ell,t} \leq R_{\text{tol}},\quad \forall t \in \mathcal{H}. 
 \label{eq:WFPI_Nmk}
\end{align}

With the introduction of $\xi_{\ell}$ which explicitly captures the wildfire ignition due a particular line $\ell$, the line outage decisions $z_{\ell}^{-}$ can be used to optimize the total wildfire risk of the network. 

To capture the influence of the de-energization decisions on the probabilities, we define a new family of probability measures $\mathbb{P}(\bm{z^{-}_{\ell}}):\mathcal{F} \rightarrow [0,1]$, where $\bm{z^{-}_{\ell}}$ is the ordered de-energization decision vector that influences the probabilities. Hence, each choice of $z_{\ell}$, induces a probability space $\biggl(\mathcal{S}, \mathcal{F}, \mathbb{P}(\bm{z^{-}_{\ell}})\biggr)$ on the same underlying sample space and the same set of events, only the probability measures are different. The probability $\pi_{\ell,s}(z_{\ell}^-)$  that a line $\ell$ with line outage decision variable $z_{\ell}^-$ in scenario $s \in \mathcal{S}$ survives (i.e. $\xi_{\ell,s}=1$) wildfire is given by,
\begin{align}
    \pi_{\ell,s}(z_{\ell}^-) = z_{\ell}^-(1 - \pi_{\ell}^1) + (1 - z_{\ell}^-)(1 - \pi_{\ell}^0), 
\end{align}
where it is assumed that there is no risk of wildfire if the line is shut-off ($z_{\ell}^-=0$). We assume each transmission line has a baseline probability of ignition $\pi_{\ell}^0$ and the PSPS line-de-energization ($z^{-}_{\ell}=0$) decision switches the failure probability to $\pi_{\ell}^1$. Similarly, the probability of wildfire due to line $\ell$ is $1 - \pi_{\ell,s}(z_{\ell}^-)$ which can be explicitly written as,
\begin{align}
    1 - \pi_{\ell,s}(z_{\ell}^-) = z_{\ell}^-\pi_{\ell}^1 + (1 - z_{\ell}^-)\pi_{\ell}^0. 
\end{align}
The negative superscript in $z^{-}_{\ell}$ is meant to signify that the de-energization decision is made before the realization of the line survival status $\xi_{\ell}$. Adding a level of uncertainty to line survival status improves the wildfire risk modeling from \cite{Greenough}, in which de-energization decisions are instantaneous due to the absence of uncertainty in wildfire ignition. The possible fire instance caused by damaged lines can be tracked with,
\begin{align} 
z^{\textrm{dn}}_{\ell}= z^{-}_{\ell}(1-\xi_{\ell}).
\label{eq:zdn}
\end{align}
For each scenario $s \in \mathcal{S}$, let $\mathcal{K}^c$ be the set of $k$ lines $\ell$ that are damaged due to wildfire and $\mathcal{K}_s^c$ its complement such that $\mathcal{L} = \mathcal{K}_s \cup \mathcal{K}_s^c$. 
The degraded state of the system due to a specific combination of $k$ line failures at time $t$ is obtained as the product of probabilities corresponding to mutually independent events as,
\begin{align}
&\pi_{s}(\bm{z^{-}_{ij}})= \prod_{\ell \in \mathcal{K}_s} 
\left(\left(z^{-}_{\ell}\right) \pi_{\ell}^0+\left(1-z^{-}_{\ell}\right) \pi_{\ell}^1\right) \nonumber \\ 
& \prod_{\ell \in \mathcal{K}_s^c}\left(\left(z^{-}_{\ell}\right)\left(1-\pi_{\ell}^0\right)+(1-z^{-}_{\ell})\left(1-\pi_{\ell}^1\right)\right).
\label{eq:NmklinefailuresDD}
\end{align}

For each scenario $s \in \mathcal{S}$, we represent the transmission line $\ell$ surviving scenario $s$ with $\xi_{\ell,s} = 1$ and conversely failing scenario, $s$ with $\xi_{\ell,s} = 0$. 
The decision-dependent scenario probabilities are given by \eqref{eq:NmklinefailuresDD}.

It is assumed that $\pi_{\ell}^0\leq \pi_{\ell}^1 \: \forall \ell \in \mathcal{L}$. Given perfect PSPS decisions, $\pi_{\ell}^1=1$, and there is no chance of a wildfire impact from following a de-energization. This simplifies \eqref{eq:NmklinefailuresDD} to:

\begin{align}
\pi_{s}(\bm{z^{-}_{ij}})= \prod_{\ell \in \mathcal{L}: \xi_{\ell,s}=0} 
&\left[\left(z^{-}_{\ell}\right) \pi_{\ell}^0+\left(1-z^{-}_{\ell}\right) \right]  \nonumber \\ 
& \quad \ldots \prod_{\ell \in \mathcal{L}: \xi_{\ell,s}=1}\left[\left(z^{-}_{\ell}\right)\left(1-\pi_{\ell}^0\right)\right].
\label{eq:NmklinefailuresDD2}
\end{align}



\begin{align}
\pi_s(\bm{z^{-}_{ij}})=\mathbb{P}(\bm{z^{-}_{ij}})(\{\bm{\xi}_{\bm{ij},s}=\bm{\tilde{\xi}}_{\bm{ij},s}\})=
    \prod_{\ell\in \mathcal{L}} \mathbb{P}(z^{-}_{\ell})(\{\xi_{\ell,s}=\tilde{\xi}_{\ell,s}\})
    \label{eq:xiscenDef}
\end{align}
In the authors' earlier work, \cite{Greenough}, the optimization was deterministic with respect to wildfire risk because there was only one scenario: the scenario that all non-zero risk (NZR) active lines eventually become damaged at the end of optimization horizon  ($\xi_{\ell}=0, R_{\ell}=\tilde{I}_{\ell} \quad \forall \ell \in \mathcal{L}$). By construction, \eqref{eq:WFPI_Nmk} also assumes that the operator has perfect foresight of the risk for each active line. However, this deterministic wildfire risk minimization is overly conservative because the scenario that all NZR active lines lead to wildfires is the rarest of all possible scenarios (see Table \ref{table:scenIEEE24DRO}).

\subsubsection{Distribution Shaping for Objective}
\label{subsubsection:Distribution Shaping for Second Stage Objective}

In our day-ahead risk-neutral PSPS formulation from \cite{Greenough}, we minimize the expected day-ahead cost given a scenario distribution $\mathbb{P}$ for the different outage scenarios $s \in \mathcal{S}$. Adding an additional layer of complexity with the de-energization dependent probability distribution ($\mathbb{P}(\bm{z^{-}_{ij}})$), we denote expected costs to be the following: 
$\mathbb{E}_{\mathbb{P}(\bm{z^{-}_{ij}})}[\bm{\Phi_{\text{DA}}}]=\sum_{s \in \mathcal{S}} \pi_{s}(\bm{z^{-}_{ij}}) \Phi_{\text{DA},s}$. 


It was shown in \cite{Greenough} that the first stage commitment costs ($f^{\text{uc}}$) are invariant of the uncertainty and for this section only we only consider the second stage operational costs $\bm{\Phi_s}$. The computational complexity of the objective function 
is highly influenced by two aspects. First, the decision dependent joint probability in Eq. \eqref{eq:NmklinefailuresDD} is a non-linear function of decision variables. Second, the number of scenarios is exponential in the number of shut-off decisions (i.e $|\mathcal{S}|=2^{|\mathcal{L}|}$). Distribution shaping, proposed in \cite{Laumanns}, provides an efficient exact method that enables one to characterize the decision-dependent scenario probabilities as a set of linear constraints. This reduces non-linearity of the objective term. We leave methods to reduce the number of scenarios through techniques, such as scenario bundling \cite{Laumanns} for future work.

Given a decision vector $\bm{z^{-}_{ij}}$, we introduce for all $\ell \in \mathcal{L}$ the truncated vector $\bm{\check{z}}_{\bm{ij},\ell}$ given by $\check{z}_{ij,k} = z_{ij,k}$ if $1 \leq k \leq \ell$, and by $\check{z}_{ij,k}= 0$ if $\ell < k \leq L$. Let us denote the corresponding scenario probabilities by $\check{\pi}_{\ell,s} = \pi_s(\bm{\check{z}}_{\bm{ij},\ell})$, and note that the trivial equality $\bm{z^{-}_{ij}} = \bm{\check{z}}_{\bm{ij},L}$ implies $\pi_s(\bm{z^{-}_{ij}}) = \pi_{L,s}$ for all $s \in \mathcal{S}$. Distribution shaping uses Bayes’ rule to construct constraints that express the linear relationship in the probability measures defined by successive truncations $\bm{\check{z}}_{\bm{ij},\ell-1}$ and $\bm{\check{z}}_{\bm{ij},\ell}$:
\begin{align}
&\min_{\bm{z_{\ell}},\bm{p_{g}},\bm{x_{d}}} \qquad  \sum_{s \in \mathcal{S}} \pi_{L,s} \Phi_{\text{DA},s}  \label{obj:SPSPS2DS}\\
&\text { s.t.} \nonumber \\
&\check{\pi}_{\ell,s} \leq \frac{\pi_{\ell}^1}{\pi_{\ell}^0} \check{\pi}_{\ell-1,s}+z^{-}_{\ell}, \forall \ell \in \mathcal{L}, s \in \mathcal{S}: \xi_{\ell,s}=0  \label{eq:ScaleUp}\\
& \check{\pi}_{\ell,s} \leq \frac{1-\pi_{\ell}^1}{1-\pi_{\ell}^0} \check{\pi}_{\ell-1,s}+z^{-}_{\ell}, \forall \ell \in \mathcal{L}, s \in \mathcal{S}: \xi_{\ell,s}=1 \label{eq:ScaleDown}\\
& \check{\pi}_{\ell,s} \leq \check{\pi}_{\ell-1,s}+1-z^{-}_{\ell}, \forall \ell \in \mathcal{L}, s \in \mathcal{S} \label{eq:NoScale}\\
& \sum_{s \in \mathcal{S}} \check{\pi}_{\ell,s}=1, \ell \in \mathcal{L} \label{eq:pisldef} \\
&\text{Operational Constraints: } \, \eqref{eq:Pg}-\eqref{eq:PowerBalance} \nonumber \\
& \bm{\check{\pi}_{\ell,s}} \in[0,1]^{|\mathcal{S}| \times  \nonumber |\mathcal{L}|}
\end{align}

\noindent where $\pi_{0,s}=\prod_{\ell \in \mathcal{L}: \xi_{\ell,s}=0} \pi_{\ell}^0 \prod_{\ell \in \mathcal{L}: \xi_{\ell,s}=1}\left(1-\pi_{\ell}^0\right)$ denotes the baseline probability of scenario $s \in \mathcal{S}$. Due to the distribution shaping relations \eqref{eq:ScaleUp}-\eqref{eq:pisldef}, the defining equalities $\check{\pi}_{\ell,s}=\pi_{s}\left(\check{z}^{-}_{\ell}\right)$ will hold for all $\ell \in \mathcal{L}, s \in \mathcal{S}$.
\subsection{Distributionally Robust Day-ahead PSPS formulation}
\label{subsubsection:Distributionally Robust Day-ahead PSPS formulation}

We use the DR-PSPS formulation for the DA stage of the PSPS optimization. A drawback of the traditional two-stage stochastic optimization 
is the assumption of perfect knowledge of the distribution of the wildfire outage probabilities, $\bm{\pi_{ij}}$, when in fact there can be tremendous uncertainty in these probabilities. To robustify our optimization problem against the uncertainty in the outage scenario probabilities, we define a distributionally robust version of the two-stage problem from the authors' earlier work \cite{Greenough}. 
 
To reduce the computational complexity of the model, we assume that the wildfire ignition is the only source of uncertainty. The formulation differs from the stochastic PSPS problem in \cite{Greenough} by assuming the uncertainty in demand is captured in a single scenario that represents the expected demand (i.e. $\mathbb{E}[\bm{p_{d,\omega}}]$); the expected WIP (i.e. $\mathbb{E}[\bm{\pi_{ij,\omega}}]$) represents the baseline WIP ($\pi^0_{\ell}$) from \eqref{eq:NmklinefailuresDD},\eqref{eq:NmklinefailuresDD2}. The expected demand and wildfire ignition are calculated as $\mathbb{E}[\bm{p_{d,\omega}}]= \sum_{\omega \in \Omega} \pi_{\omega}\bm{p}_{\bm{d},\omega} $ and $\mathbb{E}[\bm{\pi_{ij,\omega}}]= \sum_{\omega \in \Omega} \pi_{\omega}\bm{\pi}_{\bm{ij},\omega} $ respectively. The $s$ subscripts in the decision variables indicate different second-stage decisions for each outage scenario $s \in \mathcal{S}$. $\pi_{L,s}$ approximates the probability of each outage scenario $\pi_s(\bm{z^{-}_{ij}})$.

\subsubsection{DR-PSPS Objective Function}
The objective function $\Pi_{\text{DA}}$ consists of unit commitment costs ($f^{\text{uc}}$), operational constraints ($f^{\text{oc}}$), and value of lost load cost ($f^{\text{VoLL}}$) as,
\begin{align}
    &\max_{\mathbb{Q}\in\mathcal{P}}\mathbb{E}_{\mathbb{Q}}[\Pi_{\text{DA}}] =  f^{\text{uc}}(\bm{z^{\text{up}}_{g}},\bm{z^{\text{dn}}_{g}})+\max_{\mathbb{Q}\in\mathcal{P}}\mathbb{E}_{\mathbb{Q}}[\bm{\Phi_{s}}]  \label{obj:PSPS} \\
    &{f^{\text{uc}}}(\bm{z^{\text{up}}_{g}},\bm{z^{\text{dn}}_{g}}) = \mathop \sum \limits_{t \in \mathcal{H}} (\mathop \sum \limits_{g \in \mathcal{G}} {C_g^{\text{up}}{z^{\text{up}}_{g,t}} + C_g^{\text{dn}}{z^{\text{dn}}_{g,t}}} ), \label{eq:STOUC}
\end{align}
where $\bm{\Phi_s}$ is defined as,
\begin{align}
    &\bm{\Phi_s} = f_{s}^{\text{oc}}(\bm{p}_{\bm{g}, s}) + f_{s}^{\text{VoLL}}(\bm{x}_{\bm{d},s},\bm{p}_{\bm{d},s}) \\ 
    &f^{\text{oc}}(\bm{p_{g,s}}) = \mathop \sum \limits_{t \in \mathcal{H}} (\sum \limits_{g \in \mathcal{G}} {C_g p_{g,t,s}} ), \label{eq:STOOC} \\
    &{f^{\text{VoLL}}}(\bm{x_{d,s}}, \mathbb{E}_{\omega}[\bm{p_{d,\omega}}]) = \mathop \sum \limits_{t \in \mathcal{H}} (\mathop \sum \limits_{d \in \mathcal{{D}}} C_{d}^{\text{VoLL}}\left(1-x_{d,t,s}\right)\mathbb{E}_{\omega}[p_{d,t,\bm{\omega}}]). \label{eq:STOVoLL}
\end{align}
In~\eqref{obj:PSPS}-\eqref{eq:STOVoLL}, $C_g^{\text{up}}$, $C_g^{\text{dn}}$, and $C_g$ are the generator start up cost, shut down cost, and marginal cost respectively. The fraction of the load served is $x_{d,t,s} \in [0,1]$ and $C_d^{\text{VoLL}}$ is the cost incurred as a result of shedding $(1-x_{d,t,s})$ proportion of the load, $p_{d,t,s}$, 
which together with the binary variables $z_{g,t}^{\text{up}}$ and $z_{g,t}^{\text{dn}}$ capture the one-time cost incurred when bringing a generator online or offline. Let $p_{g,s} = (p_{g,1,s}, \dots, p_{g,H,s})^\top$ be the vector of $p_{g,t,s}$ for generator $g$ given scenario $s$, then $\bm{p}_{\bm{g},s}$ for all generators are denoted as $\bm{p}_{\bm{g},s} = (p_{1,s}, \dots, p_{G,s})^\top$. The variables $\bm{p}_{\bm{d},s}$, $\bm{x}_{\bm{d},s}$, $\bm{z_g}^{\text{up}}$,  $\bm{z_g}^{\text{dn}}$ are defined in a similar manner.




\subsubsection{First Stage Formulation DRO}
\label{subsubsection:First Stage Formulation DRO}
One can think of this problem as a game between the system operator who chooses to minimize economic costs based on data on outage scenarios $\bm{\tilde{\xi}_{ij}}$ from $\mathbb{P}(\bm{z^{-}_{ij}})$, while an adversary attempts to choose true probability distribution $\mathbb{Q}$ for the true outages $\bm{\xi_{ij}}$. The adversary hopes to construct an ambiguity set $\mathcal{P}$ large enough to contain the true distribution.  
\begin{align}
&\min_{\bm{z^{\text{up}}_{g}},\bm{z^{\text{dn}}_{g}}} \qquad f^{\text{uc}}(\bm{z^{\text{up}}_{g}},\bm{z^{\text{dn}}_{g}})+\max_{\mathbb{Q}\in\mathcal{P}}\mathbb{E}_{\mathbb{Q}}[\bm{\Phi_{m}}] \label{obj:SPSPS1DRO}\\
&\text { s.t. } \text{Wildfire Risk Budget: } 
\eqref{eq:WFPI_Nmk},  \nonumber \\ 
&z_{g,t} \geq \sum_{t^{\prime} \geq t-t^{\text{MinUp}}_{g}}^{t} z^{\text{up}}_{g, t^{\prime}}, \label{eq:UC1} \\
&1-z_{g,t} \geq\sum_{t^{\prime} \geq t-t^{\text{MinDn}}_{g}} z^{\text{dn}}_{g, t^{\prime}}, \label{eq:UC2}\\
&z_{g, t+1}-z_{g, t}= z^{\text{up}}_{g, t+1}-z^{\text{dn}}_{g, t+1}, \label{eq:UC3}\\
&\forall t \in \mathcal{H}, g\in\mathcal{G}, \ell \in \mathcal{L} \nonumber 
\end{align}

By definition, $\mathbb{E}_{\mathbb{Q}}[\bm{\Phi_{m}}]=\sum^{M}_{m=1} q_m\Phi_m$ in  Eq. \ref{obj:SPSPS1DRO}.

The unit commitment constraints \eqref{eq:UC1}-\eqref{eq:UC2} are used to enforce minimum up time ($t^{\text{MinUp}}_{g}$) and down time ($t^{\text{MinDn}}_{g}$) of generators. Similarly, the constraint~\eqref{eq:UC3} guarantees consistency between the binary variables $z^{\text{up}}_{g,t}$ and $z^{\text{dn}}_{g,t}$.

\subsubsection{Second Stage Formulation}
\label{subsubsection:Second Stage Formulation}
In the second stage, 
the operator solves an economic dispatch problem for each scenario:
\begin{align}
&\min_{\bm{p}_{\bm{g},s},\bm{x}_{\bm{d},s}} \quad \Phi_{s}\label{obj:SPSPS2a}\\
&\text { s.t. } \text{Distribution Shaping: } \, \eqref{eq:ScaleUp}-\eqref{eq:pisldef} \nonumber \\
&z_{g,t} \underline{p}_{g} \le p_{g,t,s} \le z_{g,t} \overline{p}_{g}, \label{eq:Pg}\\
&p^{\text{aux}}_{g,t,s}=p_{g,t,s}-\overline{p}_{g} z_{g, t}, \label{eq:RampAUX} \\
&\underline{U}_{g} \leq p^{\text{aux}}_{g, t+1,s}-p^{\text{aux}}_{g,t,s} \leq \overline{U}_{g,} \label{eq:RampNEW}\\
&p_{\ell, t,s} \leq-B_{\ell}\left(\theta_{i,t,s}-\theta_{j,t,s}+\overline{\theta}\left(1-\xi_{\ell,t,s}\right)\right), \label{eq:MaxPowerFlow} \\
&p_{\ell,t,s} \geq-B_{\ell}(\theta_{i,t,s}-\theta_{j,t,s}+\underline{\theta}\left(1-\xi_{\ell,t,s}\right)), \label{eq:MinPowerFlow}\\
&\underline{p}_{\ell,t} \, \xi_{\ell,t,s} \leq p_{\ell,t,s} \leq \overline{p}_{\ell,t}\, \xi_{\ell,t,s}, \label{eq:ThermalLimit}\\
&\sum_{g \in \mathcal{G}_{i}} p_{g,t,s}+\sum_{\ell \in \mathcal{L}_i} p_{\ell, t,s} -\sum_{d \in \mathcal{D}_{i}} x_{d,t,s} \mathbb{E}_{\omega}[p_{d,t,\bm{\omega}}]=0,\label{eq:PowerBalance}\\
&\quad \forall t \in \mathcal{H} ,\: i \in \mathcal{B},\: j \in \mathcal{N}_i ,\: \ell \in \mathcal{L}_i,\: \forall s \in \mathcal{S} \nonumber
\end{align}
where $\bm{z^{\text{up}}_{g}},\bm{z^{\text{dn}}_{g}}, \bm{z^{-}_{ij}}$, 
are obtained from the first stage.

The power of generator $g$ is limited between its minimum ($\underline{p}_{g}$) and maximum ($\overline{p}_{g}$) power generation limits in~\eqref{eq:Pg}. Similarly,~\eqref{eq:RampAUX}-\eqref{eq:RampNEW} implements the generator ramp rate limit between minimum ($\underline{U}_{g}$) and maximum ($\overline{U}_{g}$) limits. The auxiliary variable {$p^{\text{aux}}_{g, t}$} is introduced to prevent ramp violations during the startup process. To model power flow through the transmission lines, the DC-OPF approximation is used in~\eqref{eq:MaxPowerFlow}-\eqref{eq:ThermalLimit} with minimum and ($\underline{p}_{\ell,t}$) maximum ($\overline{p}_{\ell,t}$) line thermal limits. Finally, the bus power balance at each $t\in \mathcal{H}$ and every bus $i \in \mathcal{B}$ is given by the equality constraint~\eqref{eq:PowerBalance}. The generation of the DA scenarios $\omega \in \Omega$ including a total demand and WIPs for every transmission line is further discussed in Section~\ref{subsection:Data and Test Case Description}.

\subsubsection{Dual of Second Stage Formulation DRO}
\label{subsubsection:Second Stage Formulation DRO}
By applying the Kantorovich-Rubinstein Theorem, one can rewrite the worst-case second stage optimization as the following optimization that constructs a discrete joint distribution between $\mathbb{P}$ and $\mathbb{Q}$. 
\begin{align}
\max _{\bm{\gamma_{ks}} \geq 0,\bm{z^{-}_{ij}},\bm{p_{g}},\bm{x_{d}}} & \sum_{s=1}^S \sum_{m=1}^M \gamma_{s m} \Phi_m\left(\bm{z^{-}_{ij}},\bm{\xi}_{\bm{ij},m}\right) \label{obj:SPSPS2DRODef} \\
\text { s.t. } & \sum_{s=1}^S \sum_{m=1}^M \gamma_{sm } \delta_{sm}\leq \kappa \label{eq:SPSPS2DRORadius}\\
& \sum_{m=1}^M \gamma_{s m}=\pi_s(\bm{z^{-}_{ij}}), \label{eq:SPSPS2DROMargP}
s \in\{1, \ldots, S\} \\
& \sum_{s=1}^S \gamma_{s m}=q_m, \label{eq:SPSPS2DROMargQ}
m \in\{1, \ldots, M\} \\
&\text{Distribution Shaping: } \, \eqref{eq:ScaleUp}-\eqref{eq:pisldef} \nonumber \\
&\text{Operational Constraints: } \, \eqref{eq:Pg}-\eqref{eq:PowerBalance} \nonumber
\end{align}

\noindent where $\kappa$ controls the distance or dissimilarity between outages scenarios in the ambiguity set and the sample outages. $\delta_{sm}=\delta\left(\bm{\xi}_{\bm{ij},m}-\bm{\tilde{\xi}}_{\bm{ij},s}\right)$ is a function that evaluates a binary total variation distance between training samples outage samples $\bm{\tilde{\xi}}_{\bm{ij},s}$ and true outages $\bm{\xi}_{\bm{ij},m}$.  
\begin{align}
\delta_{s m}=
    \begin{cases}
      0, & \text{if}\ \bm{\tilde{\xi}}_{\bm{ij},s}=\bm{\xi}_{\bm{ij},m} \\
      1, & \text{otherwise}
    \end{cases}
    \label{eq:TVDistance}
\end{align}
We choose our distance metric $\delta_{sm}$ to be the total variational distance to reduce the computational complexity of the optimization. We leave the incorporation of other metrics, such as, the 1-norm or 2-norm Wasserstein distance for future work \cite{Esfahani}. It should be noted that the original risk-neutral version of the DR-PSPS problem (or the risk neutral SPSPS from \cite{Greenough}) can be recovered when setting $\kappa=0$. This would force $\gamma_{sm}=0 \: \forall s\neq m$. Then, the marginal probabilities would be equal (e.g. $q_m=\pi_s(\bm{z^{-}_{ij}})$ for each respective $s,m$ sample) and the DRO and risk-neutral stochastic objectives would be equal (e.g. $\mathbb{E}_{\mathbb{P}(\bm{z^{-}_{ij}})}[\bm{\Phi_{\text{s}}}]=\max_{\mathbb{Q}\in\mathcal{P}}\mathbb{E}_{\mathbb{Q}}[\bm{\Phi_{s}}]$).

Next, we dualize \eqref{obj:SPSPS2DRODef}-\eqref{eq:SPSPS2DROMargP} to create the following quadratic program:
\begin{align}
& \min _{\tau \geq 0, \bm{v},\bm{z^{-}_{ij}},\bm{p_{g}},\bm{x_{d}}} \tau \kappa+\sum_{s \in \mathcal{S}} \pi_s(\bm{z^{-}_{ij}}) v_s  \label{obj:SPSPS2DRODual}\\
& \text { s.t. } v_s+\tau\delta_{sm} \geq \Phi_m\left(\bm{z^{-}_{ij}},\bm{\xi}_{\bm{ij},m}\right), \label{eq:SPSPS2DRODual} \\
&\text{Distribution Shaping: } \, \eqref{eq:ScaleUp}-\eqref{eq:pisldef} \nonumber \\
&\text{Operational Constraints: } \, \eqref{eq:Pg}-\eqref{eq:PowerBalance} \nonumber \\
&\forall s \in\{1, \ldots, S\}, m \in\{1, \ldots, M\} \nonumber
\end{align}

\subsubsection{Second Stage Formulation DRO MILP Approximation}
\label{subsubsection:Second Stage Formulation DRO MILP Approximation}

Using the method in \cite{Noyan} we input the solution to $\tau$. Let us introduce $\forall j \in \mathcal{S}$ a binary variable $\beta_j$ that takes the value 1 if and only if the nominal value-at-risk level $\kappa$ of $\bm{\Phi}$ is equal to the realization $\Phi_j$. Then, recalling condition $\tau=\max(\bm{\Phi})-\text{VaR}_{\kappa}(\bm{\Phi})$, at an optimal solution of \eqref{obj:SPSPS2DRODual} - \eqref{eq:SPSPS2DRODual} we have $v_{s} =\max{\{\Phi_s,\max{(\bm{\Phi})}-\tau}\}=\max\{{\Phi_s, \sum_{j \in \mathcal{S}} \Phi_j\beta_j\}}= \sum_{j \in \mathcal{S}} \max{\{\Phi_s,\Phi_j\}}\beta_j$. We can now use McCormick envelopes to linearize the quadratic terms $\pi_{L,s}v_s$ in the objective \eqref{obj:SPSPS2DRODual} by introducing the auxiliary variables $w_{sj}=\pi_{L,s}\beta_j$, which leads to the following MILP formulation of
\cite{Noyan}:
\begin{align}
\min_{\bm{w_{sj}},\bm{\beta_{j}},\bm{z^{-}_{ij}},\bm{p_{g}},\bm{x_{d}}} & \sum_{s \in \mathcal{S}} \sum_{j \in \mathcal{S}} \max\{\Phi_s, \Phi_j\} w_{s j} + \kappa \tau  
\label{eq:SPSPS2DROMILP} \\
&\text { s.t. } 
\tau=\max{(\bm{\Phi})}-\sum_{j \in \mathcal{S}} \Phi_j \beta_j \label{eq:SPSPS2DROMCtau}\\
& w_{s j} \leq \pi_{L,s}, \label{eq:SPSPS2DROMCpiL}\\
& w_{s j} \leq \beta_j, \label{eq:SPSPS2DROMCbeta}\\
& w_{s j} \geq \pi_{L,s}+\beta_j-1, \label{eq:SPSPS2DROMCAll}\\
& \sum_{j \in \mathcal{S}} \beta_j=1, \label{eq:SPSPS2DROBetaDef}\\
&\text{Distribution Shaping: } \, \eqref{eq:ScaleUp}-\eqref{eq:pisldef} \nonumber \\
&\text{Operational Constraints: } \, \eqref{eq:Pg}-\eqref{eq:PowerBalance} \nonumber \\
& \forall s, j \in \mathcal{S}, \quad \bm{\beta_{j}} \in\{0,1\}^S, \quad \bm{w_{sj}} \in[0,1]^{S \times S} \nonumber .
\end{align}

\vspace{-1em}

\subsection{Real-Time Simulation for Out-of-Sample Performance}
\label{subsubsection:Real-Time Outage Simulation}

After generator commitments ($\bm{z_g}$) and line shut-off decisions ($\bm{z^{-}_{ij}}$) are made, the system operator evaluates their performance using realized nodal demands and line outages. Instead of expected values over scenarios $\omega \in \Omega$, the operator uses realized samples ($p_{d,\omega'}$, $\xi_{l,\omega'}$) from scenarios $\omega' \in \Omega^{\text{RT}}$. WIP forecasts $\pi_{l,\omega'}$ are used to generate 200 Monte Carlo samples of likely outage scenarios, $\bm{\xi}_{\bm{ij},\omega'}$. One demand realization is selected, and uniform probabilities are assigned: $\pi_{\omega'} = \frac{1}{200}$. In practice, these realizations can also come from improved real-time forecast scenarios. The RT optimization problem is the following:
\begin{align}
&\min_{\bm{p_{g,\omega'}}, \bm{x_{d,\omega'}}}  \: \sum_{\omega'\in \Omega^{\text{RT}}} \pi_{\omega'} {\Pi }_{\omega' } 
\\
&\text { s.t. } 
\nonumber\\
&\Pi_{\omega'} = f^{\text{uc}}(\bm{z^{\text{up}}_{g}},\bm{z^{\text{dn}}_{g}})+f^{\text{oc}}(\bm{p}_{\bm{g},\omega'}) + f^{\text{VoLL}}(\bm{x}_{\bm{d},\omega'},\bm{p}_{\bm{d},\omega'}) \nonumber \\ 
& \text{Generator Capacity Bounds with $\bm{z_{g}}$: } 
 \,\eqref{eq:Pg} \nonumber \\
&\text{Generator Ramping with $\bm{z_{g}}$: } 
 \, \eqref{eq:RampAUX}-\eqref{eq:RampNEW} \nonumber\\
&\text{Optimal Power Flow with $\bm{z}^{-}_{\bm{ij}}$ \& $\bm{\xi}_{\bm{ij},\omega'}$:} \, \eqref{eq:MaxPowerFlow}-\eqref{eq:ThermalLimit} \nonumber\\
&\text{Real-time Demand Balance with $\bm{p}_{\bm{d},\omega'}$: }\, \eqref{eq:PowerBalance} \nonumber \\
& \hspace{0.5 cm} \:\; i \in \mathcal{B} ,\: \forall \omega' \in \Omega^{\text{RT}}, \: \forall t \in \mathcal{H} \nonumber 
\end{align}
Monte Carlo simulations were generated based on daily WIP forecasts for an entire month.

\section{Results}
\label{section:Results}

\subsection{Data and Test Case Description} \label{subsection:Data and Test Case Description}
\subsubsection{Grid Models Description} \label{subsubsection:Grid models description}
We use the IEEE RTS 24-bus system (Fig.~\ref{fig:IEEE24colored}) with assigned WLFP, SVI, and acres burned values. The system includes 68 generators (3 wind, 20 RTPV, 14 utility-scale PV, 4 hydro, 1 synchronous condenser, 5 CC, 15 CT, and 6 steam), 17 loads, and 38 transmission lines. Generator details, including location, capacity, and costs, follow \cite{RTSGMLC}. The model represents Region 300 of the 73-bus RTS-GMLC and is geographically placed near Los Angeles, CA. Daily regional load profiles from Region 300 are scaled to match the RTS bus-level maximum daily loads, while hourly load data are taken from \cite{RTSGMLC}.

\subsubsection{Transmission Line WLFP, SVI, and Acres Burned Assignment} \label{subsubsection:Transmission line WLFP assignment}
\begin{figure*}[t]
\centering
\subfloat[]{\includegraphics[width=6.0cm]{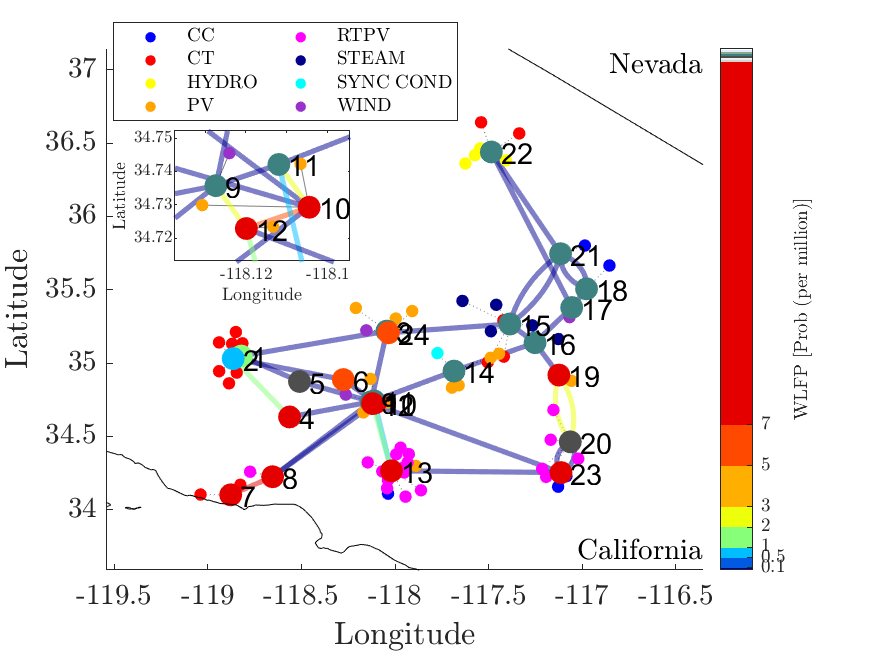}}
\subfloat[]{\includegraphics[width=6.0cm]{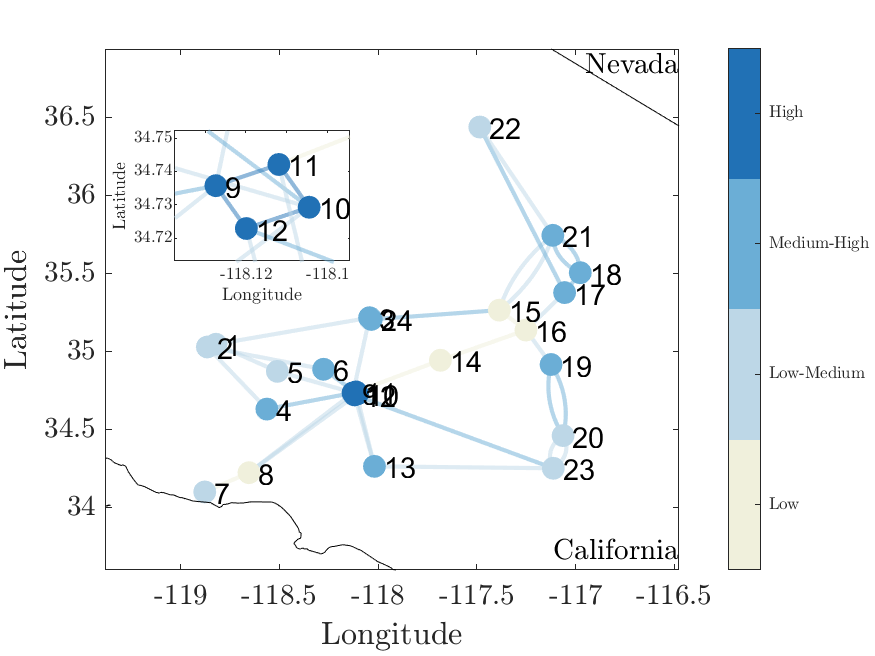}}
\subfloat[]{\includegraphics[width=6.0cm]{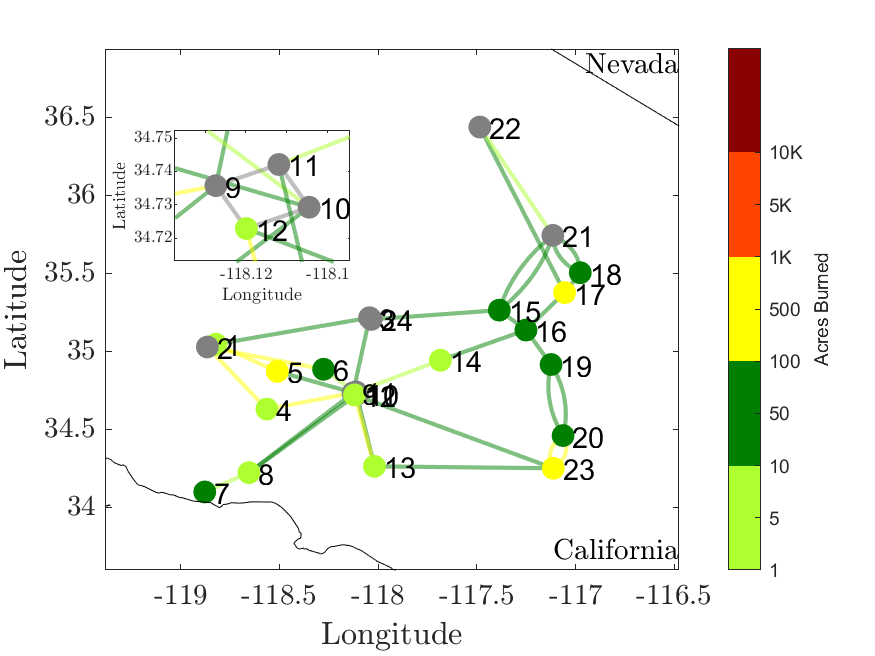}}
\caption{IEEE RTS 24-bus system schematic with each transmission line and bus highlighted to depict its WLFP (a), SVI \cite{SVI} (b), and expected acres burned from a Pyregence forecast on January 27, 2025 (c). The geographic layout is from \cite{RTSGMLC}.} 

\label{fig:IEEE24colored} 
\end{figure*}

Day-ahead WLFP forecasts are downloaded as .tiff files from \cite{WLFP}. For each bus, WLFP values are assigned by averaging the four nearest .tiff grid points based on predefined bus coordinates \cite{RTSGMLC}. Transmission line WLFP values are obtained by averaging the WLFP values along the Bresenham path between bus endpoints \cite{Bresenham}. The same method is applied to assign acres burned using hourly Pyrecast forecasts, which are then averaged daily (Fig.~\ref{fig:IEEE24colored}c). SVI data are sourced from the CDC \cite{SVI}. Each bus is assigned the SVI of its surrounding census tract. Line-level SVI is computed as the average SVI of all intersecting census tracts (Fig.~\ref{fig:IEEE24colored}b).

Simulations run over a 24-hour horizon with hourly decision intervals. Input scenarios for PSPS optimization are generated via a tree reduction algorithm from \cite{Gröwe-Kuska}, using three months of historical hourly total load and average bus-level WLFP. Unlike \cite{Greenough}, which only considers deterministic line-level WFPI, we model joint scenarios of demand and cumulative bus-level WLFP to better reflect local wildfire risks.

\begin{figure}[ht]
\centering
\vspace{-1em}
\subfloat{\includegraphics[width=4.42cm]{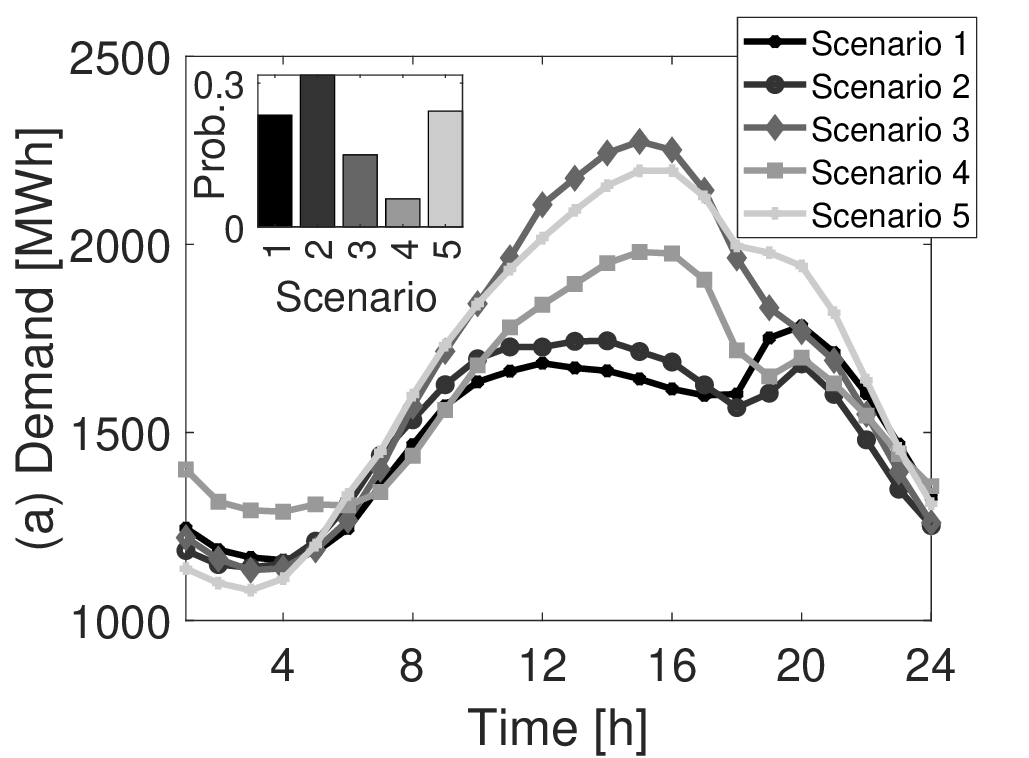}} \hfill
\subfloat{\includegraphics[width=4.42cm]{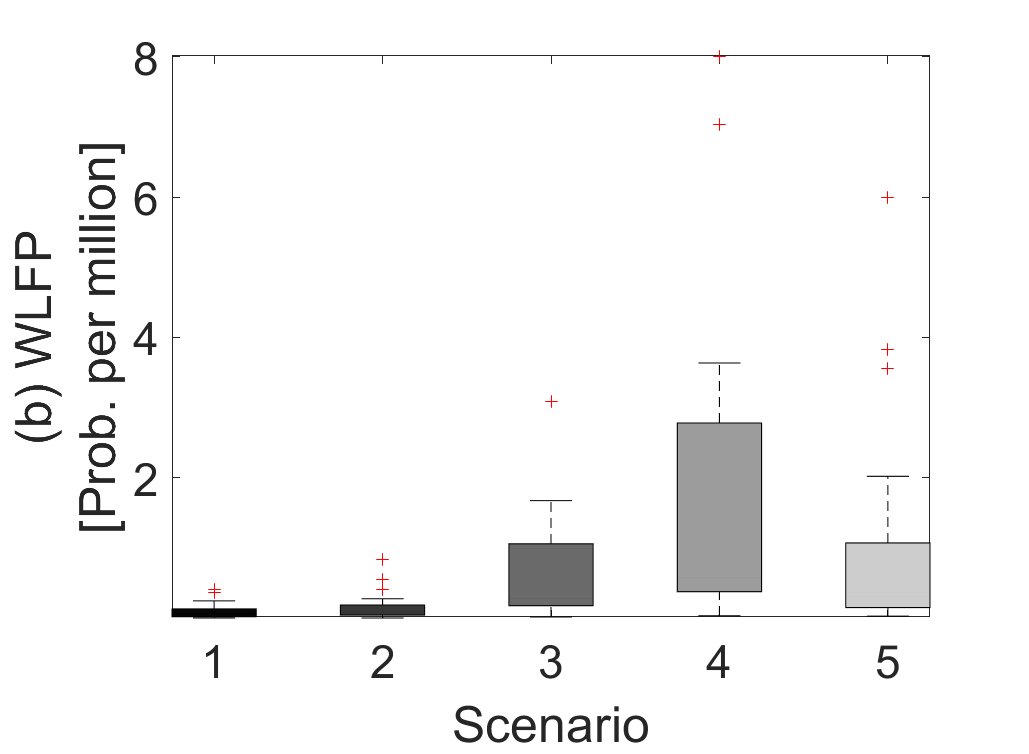}}
\caption{Five scenarios for the RTS derived from the tree reduction based on WLFP and load data from 2020. Box plots of the bus WLFP for each scenario are also shown.}
\label{fig:Load_Profiles_IEEE24_WFPI_WLFP}
\end{figure}

The five most representative 24-hour demand-WLFP scenarios are extracted and mapped to the RTS loads. For each scenario, a nearest-neighbor search identifies historical days with similar bus-level WLFP, and the corresponding line-level WLFP values are used. Fig.\ref{fig:Load_Profiles_IEEE24_WFPI_WLFP}a shows the WLFP distribution, while Fig.~\ref{fig:Load_Profiles_IEEE24_WFPI_WLFP}b provides the demand time series and scenario probabilities. WLFP values are converted to WIP using conditional means from USGS reliability diagrams \cite{WLFP}. Expected demand and WIP are used in the optimization: demand is included in the power balance constraint \eqref{eq:PowerBalance}, and WIP informs the wildfire constraints \eqref{eq:WFPI_Nmk} and \eqref{eq:ScaleUp}-\eqref{eq:pisldef}. To manage the exponential growth of scenario sets ($\sim 2^{\rm NZR}$), we focus on the top three NZR lines. The resulting outage scenarios are listed in Table~\ref{table:scenIEEE24DRO}. WLFP values are scaled by $10^4$ to ensure line outages have probabilities above 1\%, consistent with prior studies \cite{Moreira,Pianco}.

\begin{table}[ht]
\begin{center}
\caption{Table of the eight active line status scenarios for the lines of the IEEE 24 bus system with the highest WLFP}
\begin{tabular}{l m{1cm} m{1cm} m{1cm} m{2.4cm}}
Scen. \#& $\xi_{2,4}$ & $\xi_{7,8}$ & $\xi_{10,12}$ & Initial Prob: $\pi_{0,s}$ \\ 
\hline 
 (1) & 0 & 0 & 0 & $5.54\times 10^{-6}$  \\ 
 (2) & 0 & 0 & 1 & $2.44\times 10^{-4}$ \\ 
 (3) & 0 & 1 & 0 & $3.01\times 10^{-4}$\\ 
 (4) & 0 & 1 & 1 & $1.33\times 10^{-2}$\\ 
 (5) & 1 & 0 & 0 & $3.95\times 10^{-4}$ \\ 
 (6) & 1 & 0 & 1 & $1.74\times 10^{-2}$ \\ 
 (7) & 1 & 1 & 0 & $2.15\times 10^{-2}$\\ 
 (8) & 1 & 1 & 1 & 0.947\\ 
\hline 
\end{tabular}
\label{table:scenIEEE24DRO}
\end{center}
\end{table}
\vspace{-1.5em}

\subsection{IEEE 24 Bus PSPS Optimization with Equal Risk Among Transmission Lines}

In the simulations, the Value of Lost Load ($C_{d}^{\text{VoLL}}$) is set to \$5,000 per MWh for all demands \cite{Trakas2,Mohagheghi,Farzin}. Since the value of losing 1~MWh is at least one order magnitude higher than the cost to produce 1 MWh, the VoLL is typically the largest contributor to the total economic costs. 
We depict the total cost, commitments, and de-energization strategies among DR-PSPS strategies at varying levels of distributional robustness. Then, we compare those DR-PSPS strategies to the WS approach to show the cost of not having perfect information for the upcoming outages.

\begin{figure}[ht]
\centering
\includegraphics[width=\columnwidth]{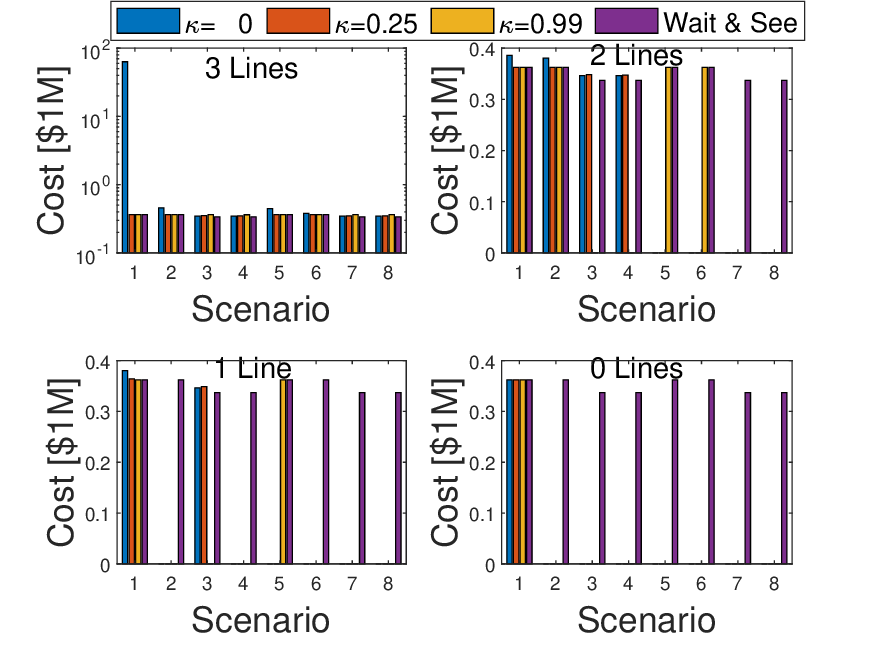}
\caption{Total cost per scenario for the IEEE 24-bus system over a 24-hour horizon under different first-stage strategies ($\kappa=0$, $0.25$, $0.99$, and Wait \& See). 
}
\vspace{-2em}
\label{fig:ScenarioCostBreakdownDROIEEE1424h}
\end{figure}

\begin{figure}[ht]
\centering
\includegraphics[width=\columnwidth]{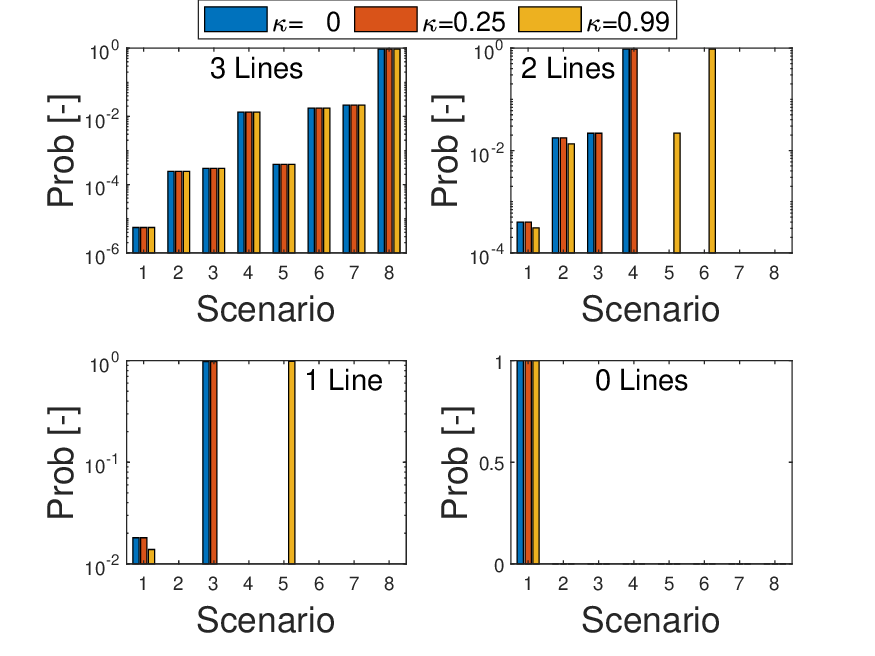}
\caption{Outage scenario probabilities for the IEEE 24-bus system over a 24-hour horizon under different first-stage DR optimization strategies ($\kappa=0$, $0.25$, $0.99$). Scenarios with $z^{-}_{\ell}=0$ have zero probability ($\pi_{s}=0$ for $\xi_{\ell,s}=1$).}
\label{fig:ScenarioProbBreakdownDROIEEE1424h}
\end{figure}

The cost per scenario breakdown between the Wait \& See approach and $\kappa=\{0,0.25,0.99\}$ are summarized for no more than 3, 2, 1 and 0 active lines in Fig. \ref{fig:ScenarioCostBreakdownDROIEEE1424h} and the probabilities of each scenario are given in Fig. \ref{fig:ScenarioProbBreakdownDROIEEE1424h}.

We propose a metric that calculates the cost difference between WS and distributionally robust approaches called the Distributionally Robustified Expected Value of Perfect Information (REVPI). Equation~\eqref{eq:maxCvar}~\cite{Noyan} shows that robustified expectation objective is equivalent  to a $\max-\text{CVaR}_{\kappa}$ objective,
\begin{align}
\max_{\mathbb{Q}\in\mathcal{P}}\mathbb{E}_{\mathbb{Q}}[\bm{\Phi_{s}}]=\kappa \max(\bm{\Phi_s})+(1-\kappa)\text{CVaR}_{\kappa}(\bm{\Phi_{s}}).
\label{eq:maxCvar}
\end{align}
Furthermore, the outer weighted sum with weighting parameter $\kappa$ is the balance between the maximum value of the total costs for each scenario (denoted as $\Pi_{s}$) among all scenarios $s \in \mathcal{S}$, and its $\kappa$-quantile expected shortfall or  $\text{CVaR}_{\kappa}(\Pi_{s})$. 

It can be shown that $\text{CVaR}_\epsilon$ is translationally invariant \cite{Noyan},
\begin{align*}
    \text{CVaR}_\kappa ( \Pi_s)=f^{\text{uc}}(\bm{z^{\text{up}}_{g}},\bm{z^{\text{dn}}_{g}})+\text{CVaR}_\kappa(\bm{\Phi_{s}}), \\
    \max( \Pi_s)=f^{\text{uc}}(\bm{z^{\text{up}}_{g}},\bm{z^{\text{dn}}_{g}})+\max(\bm{\Phi_{s}}),
\end{align*}
We modify the traditional Value of Perfect Information (VPI) metric (\cite{Noyan}) to be REVPI. 
\vspace{-0.5em}
\begin{align}
&\operatorname{REWS}=(\kappa)\operatorname{\max_{s}}{\Bigl(}\underset{\bm{z^{\text{up}}_{g}},\bm{z^{\text{dn}}_{g}}}{\min} \Pi_s(\bm{z^{\text{up}}_{g}}(\bm{\xi_s}),\bm{z^{\text{dn}}_{g}}(\bm{\xi_s}),\bm{\xi_s}){\Bigr)}+ \ldots +\nonumber \\ &\quad (1-\kappa)\operatorname{CVaR}_{\kappa}{\Bigl(}\underset{\bm{z^{\text{up}}_{g}},\bm{z^{\text{dn}}_{g}}}{\min}\Pi_s(\bm{z^{\text{up}}_{g}}(\bm{\xi_s}),\bm{z^{\text{dn}}_{g}}(\bm{\xi_s}),\bm{\xi_s}){\Bigr)} \label{eq:REWS} \\
&\operatorname{RERP}=\min_{\bm{z^{\text{up}}_{g}},\bm{z^{\text{dn}}_{g}}}\biggl\{\kappa\operatorname{max}[\bm{\Pi}(\bm{z^{\text{up}}_{g}},\bm{z^{\text{dn}}_{g}},\bm{\xi})]+ \ldots +\nonumber \biggr.\\
&\quad \biggl.(1-\kappa)\operatorname{CVaR}_{\kappa}\left(\bm{\Pi}(\bm{z^{\text{up}}_{g}},\bm{z^{\text{dn}}_{g}},\bm{\xi})\right)\biggr\} \label{eq:RERP} \\
&\operatorname{REVPI}=\operatorname{RERP}-\operatorname{REWS} \label{eq:REVPI}
\end{align}
Notice in the definition of REWS \eqref{eq:REWS}, we compute unit commitment decisions and total economic costs for each outage scenario separately, and then compute the maximum and CVaR for a total cost vector. In the RERP, which is known as DR-PSPS in \eqref{obj:SPSPS1DRO}, we have one set of unit commitment decisions for the unit commitment problem. 
As seen in Fig.~\ref{fig:ScenarioCostBreakdownDROIEEE1424h} and Table~\ref{table:REVPI24h}, there is a cost benefit to perfect insight into the outage scenarios because the commitment strategy is not constant with respect to the outage scenario. More specifically, WS commits a different number of generators than any of the distributionally robust approaches. 

\begin{table}[ht]
\begin{center}
\caption{Robustified Expected Value of Perfect Information (REVPI from \eqref{eq:REVPI} in [\$1,000]) at various active line and distributionally robust settings for the full optimization 24~hr period on the IEEE RTS 24-bus system.}
\begin{tabular}{m{1.1cm} m{1.45cm} m{1.45cm} m{1.45cm} m{1.45cm}}
 REVPI($\kappa$) & 3 Lines & 2 Lines & 1 Line & 0 Lines \\ 
\hline 
$0.999$ & 0 & 0 & 0 & 0\\ 
$0.99$ & 0 & 0 & 0 & 0 \\ 
$0.75$ & $2.44$ & $2.41$ & $2.41$ & $5.86$ \\ 
$0.50$ & $5.03$ & $5.01$ & $5.01$ & $12.2$ \\ 
$0.25$ & $7.64$ & $7.64$ & $9.11$ & $18.5$ \\ 
$0.00$ & $9.75$ & $9.36$ & $9.36$ & $24.8$ \\ 
\hline 
\end{tabular}
\label{table:REVPI24h}
\end{center}
\end{table}

\subsection{Out-of-Sample Performance}

In Table \ref{table:RealTime24h}, we tabulate total expected costs from the Monte-Carlo simulations averaged over July (i.e. 31 days during peak wildfire season). As shown in Table \ref{table:RealTime24h}, the risk-neutral approach ($\kappa=0$) often leads to the least robust to out-of-sample outages, while a higher level of distributional robustness ($\kappa=0.99$) resulted in lower out-of-sample costs. The difference in Out-of-Sample cost for the case of zero lines is purely a result of different commitment strategies leading to the same optimal solutions day-ahead.

\begin{table}[ht]
\begin{center}
\caption{Expected Total Out-of-Sample Costs in [\$1 $\times 10^5$] at various active line and DR settings for the full 24~hr period on the IEEE RTS 24-bus system.}
\begin{tabular}{m{2cm} m{1.15cm} m{1.15cm} m{1.15cm} m{1.15cm}}
Opt Value ($\kappa$) & 3 Lines & 2 Lines & 1 Lines & 0 Lines \\ 
\hline 
0.999 & 2.616 & 2.650 & 2.650 & 2.616 \\ 
0.99 & 2.623 & 2.623 & 2.616 & 2.616 \\ 
0.75 & 2.650 & 2.650 & 2.650 & 2.650 \\ 
0.50 & 2.650 & 2.650 & 2.650 & 2.650 \\ 
0.25 & 2.650 & 2.650 & 2.697 & 2.650 \\ 
0.00 & 2.660 & 2.660 & 2.660 & 2.643 \\ 
\hline 
\end{tabular}
\label{table:RealTime24h}
\end{center}
\end{table}

\vspace{-2.5em}
\subsection{Day-ahead Commitment and Operational Costs}

\begin{figure}[ht]
\centering
\subfloat[]{\includegraphics[width=4.4cm]{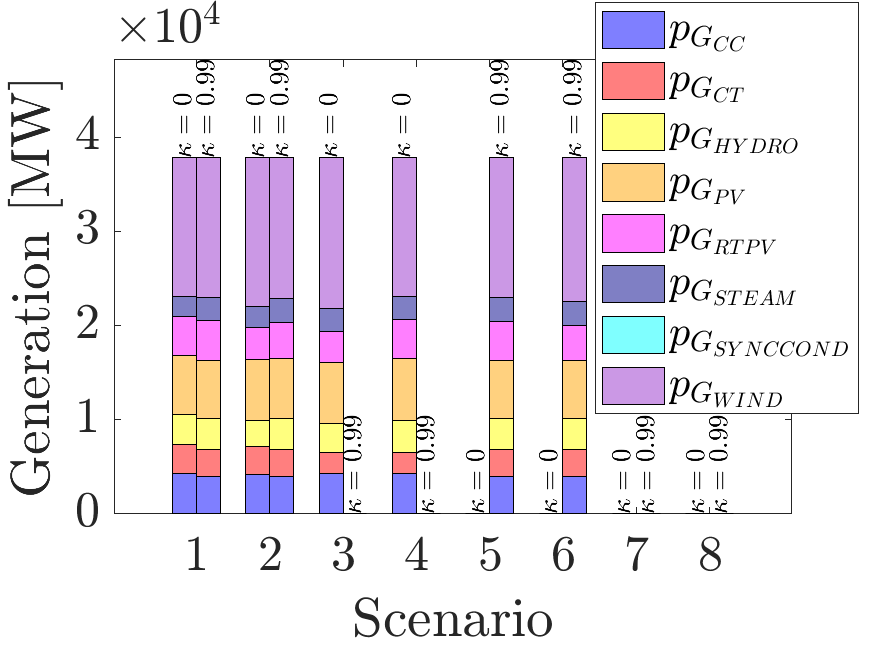}}
\hfil
\subfloat[]{\includegraphics[width=4.4cm]{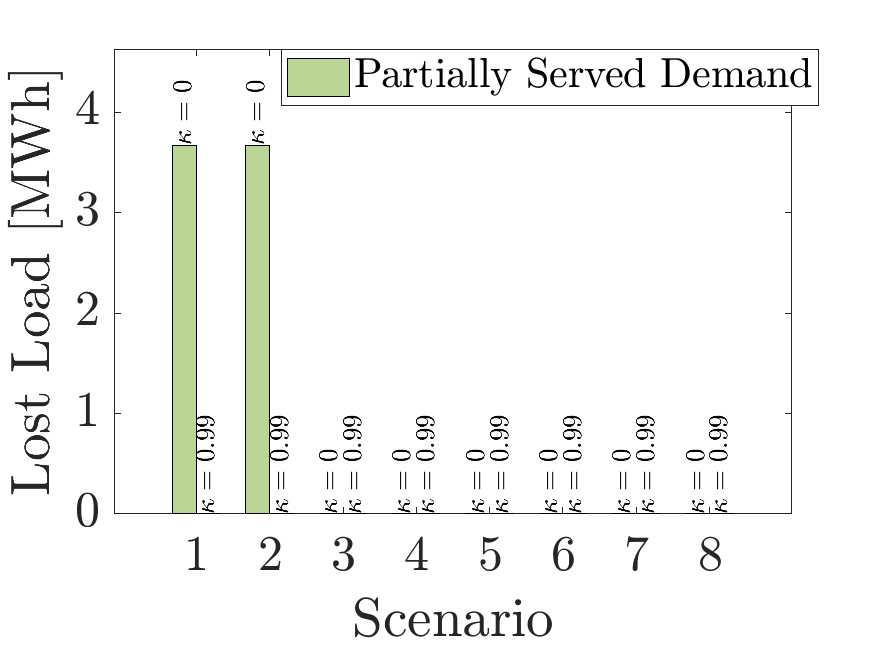}}
\hfil
\subfloat[]{\includegraphics[width=4.4cm]{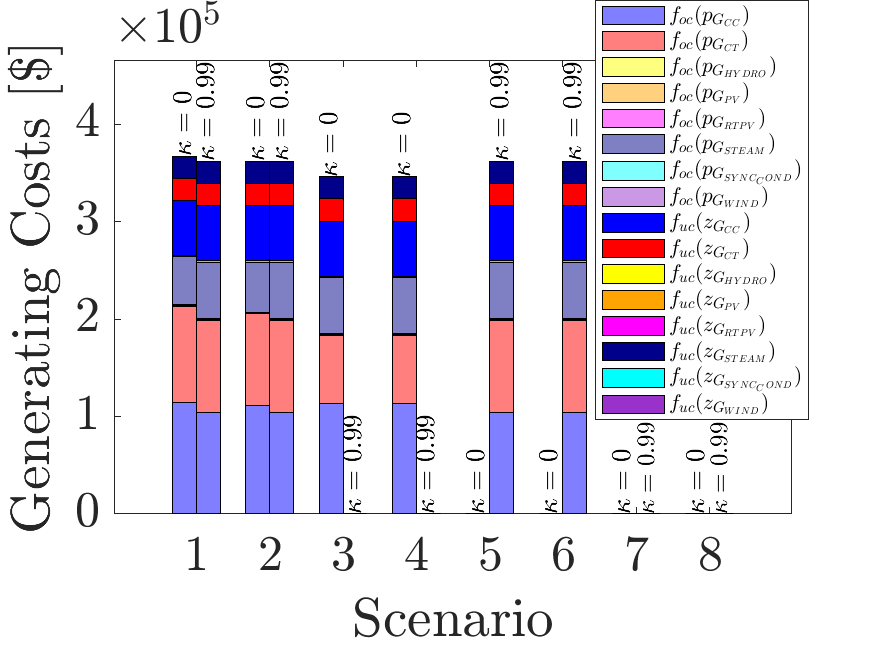}}
\hfil
\subfloat[]{\includegraphics[width=4.4cm]{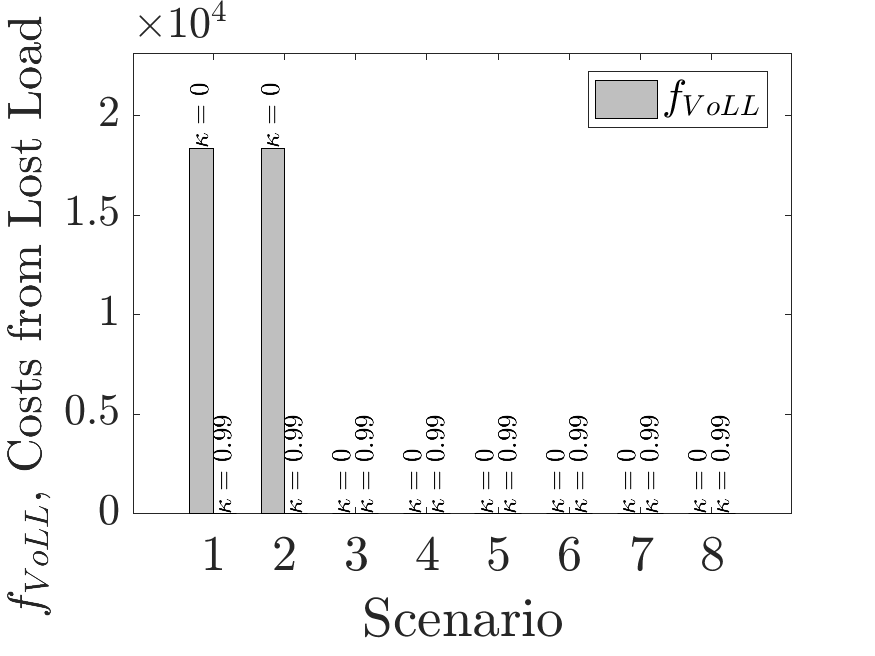}}
\caption{PSPS (a) generation, (b) load shedding, (c) total costs (excluding VoLL), and (d) VoLL costs for the IEEE RTS 24-bus system optimized over 24 hours (e.g. the case when at most 2 lines are active Fig. \ref{fig:ScenarioCostBreakdownDROIEEE1424h} (b)). Left bars are risk-neutral ($\kappa=0$), and right bars are risk-averse ($\kappa=0.99$) results.} 
\label{fig:IEEE24BBarGenCostCompareleq2L}
\end{figure}

We focus on the differences in the costs between a risk-neutral ($\kappa=0$) and a risk-averse ($\kappa=0.99$) approach for combinations of at least 1 line outage optimized over 24 hours. Figure \ref{fig:IEEE24BBarGenCostCompareleq2L}(c) shows that the risk-averse strategy incurs more operating and commitment costs in all scenarios except the least likely scenario (scenario 1). The risk-neutral strategy incurs significant load loss in the two least likely scenarios (1 \& 2), leading to the most costly decisions in the least likely scenario (scenario 1) as well as the least likely scenario among those with two damaged lines (scenario 2) 

\begin{figure}[ht]
\centering
\subfloat{\includegraphics[width=\columnwidth]{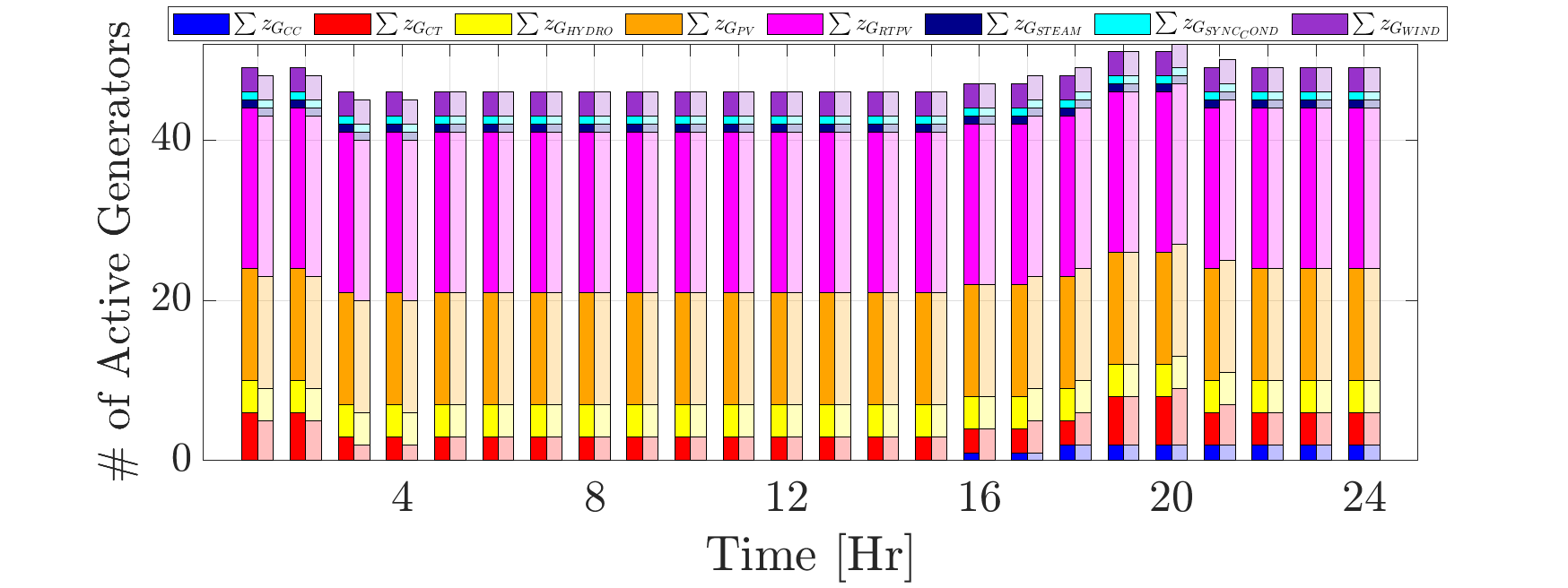}}
\hfil
\subfloat{\includegraphics[width=\columnwidth]{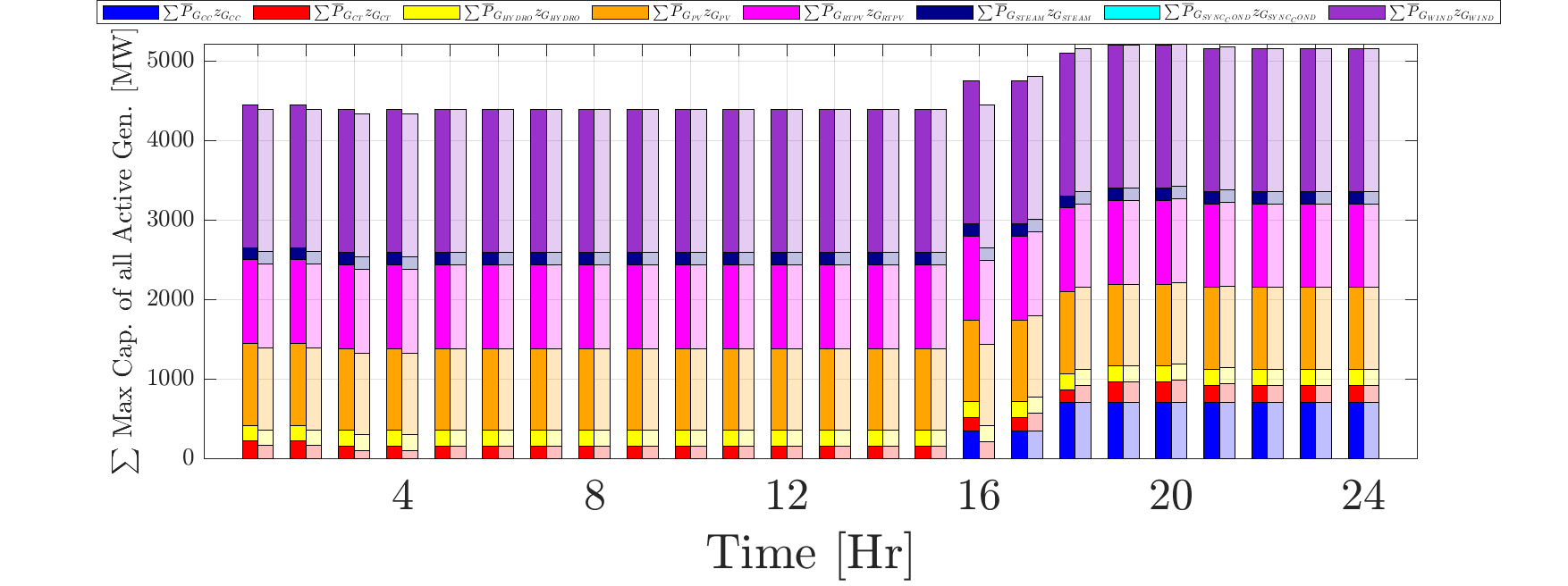}}
\caption{(Top) Active generator count; (Bottom) max available capacity. Left bars: risk-neutral ($\kappa=0$), right bars: risk-averse ($\kappa=0.99$). The risk-averse strategy commits more CT units, increasing capacity during peak hours (see Fig.~\ref{fig:IEEE24BBarGenCostCompareleq2L}).
} 
\label{fig:IEEE24CapAveCompare_bygen_1stDRO}
\end{figure}

Figure \ref{fig:IEEE24CapAveCompare_bygen_1stDRO} shows the risk-averse model commits extra CT generators at 1600~h, 1700~h, 1900~h, 2000~h 
leading to a larger maximum generation capacity 
at the expense of higher total commitment costs shown in Fig.~ \ref{fig:IEEE24BBarGenCostCompareleq2L}(c).

\vspace{-1em}
\subsection{Costs vs. WF Risk based on Acres Burned and SVI}
\begin{figure}[ht]
\centering
\includegraphics[width=\columnwidth]{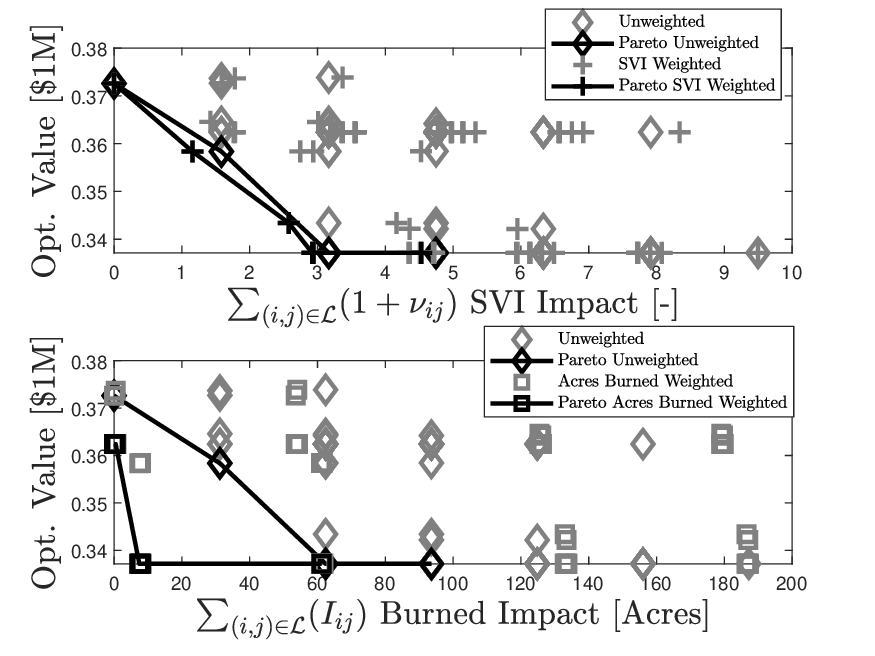}
\caption{Pareto-optimal (black) vs. sub-optimal (gray) solutions: total cost vs. wildfire risk via SVI (top) and vs. acres burned (bottom) across 64 outage scenarios. These 64 scenarios are a result of $2^6$ permutations of outages from the 6 riskiest lines of the IEEE RTS 24-bus system.} 
\label{fig:IEEESVIABNMK_Compare}
\end{figure}

In Fig. \ref{fig:IEEESVIABNMK_Compare}, we plot  
total costs versus wildfire risk for unweighted, SVI-weighted, and acres burned-weighted wildfire impact in~\eqref{eq:WFPI_Nmk}. For computational tractability of sweeping through all scenarios, we plot the results from WS. Fig.~\ref{fig:IEEESVIABNMK_Compare} shows the reduction in Pareto optimal impact by weighting each line individually. 

\section{Conclusion}
\label{section:Conclusion}
This paper shows how an operator's choice of transmission line de-energizations not only affects the scenario probabilities but also affects operational decisions. A DRO version of the SPSPS framework with de-energization-dependent scenario probabilities extends the authors' previous works \cite{Greenough, Greenough2}  and minimizes total expected economic cost through an implementation on the IEEE RTS 24-bus system in two stages.  


This study is limited to showing the impact of DR-PSPS with an ambiguity set based on the total variation distance between the nominal and candidate worst-case distribution. More advanced constructions for ambiguity sets can be considered. Additional studies may be needed to show how the PSPS optimization total economic cost would be affected by different ambiguity set constructions and wildfire forecasts developed by utilities or by other proposed methods in the literature (e.g. the wildfire forecasting methods developed in \cite{Umunnakwe, Bayani, Bayani2}).


\clearpage

\end{document}